\documentclass[journal,12pt,onecolumn,draftclsnofoot,]{IEEEtran}

\usepackage[utf8]{inputenc}
\usepackage[english]{babel}
\usepackage{epsfig}
\usepackage{amsfonts}
\usepackage{ifpdf}
\usepackage{multirow}
\usepackage{amssymb}
\usepackage{amsthm}
\usepackage{mathtools}
\usepackage{mathrsfs}
\usepackage{textgreek}
\usepackage{grffile}
\usepackage{xcolor}
\usepackage{bm}

\usepackage{cancel}
\usepackage{nccmath}
\usepackage{cuted}
\usepackage{setspace}

\usepackage[usestackEOL]{stackengine}

  \stackMath

\ifCLASSINFOpdf
\else
\fi
\usepackage{amsmath}
\usepackage{algorithmic}
\usepackage{algorithm}
\usepackage{graphicx}

\newcommand{\norm}[1]{\left\lVert#1\right\rVert}

\ifCLASSOPTIONcompsoc
  \usepackage[caption=false,font=normalsize,labelfont=sf,textfont=sf]{subfig}
\else
  \usepackage[caption=false,font=footnotesize]{subfig}
\fi

\usepackage{stfloats}
\begin{document}
\title{RIS-Aided Kinematic Analysis for Remote Rehabilitation}
\author{{Don-Roberts~Emenonye, Anik~Sarker,
 Alan~T.~Asbeck,  Harpreet~S.~Dhillon, and
R.~Michael~Buehrer
}
\thanks{D.-R. Emenonye, H. S. Dhillon and R. M.  Buehrer are with Wireless@VT,  Bradley Department of Electrical and Computer Engineering, Virginia Tech,  Blacksburg,
VA, 24061, USA. Email: \{donroberts, hdhillon, rbuehrer\}@vt.edu. } \thanks{ Anik Sarker and  Alan T. Asbeck are with the Department of Mechanical Engineering, Virginia Tech,  Blacksburg,
VA, 24061, USA. Email: \{aniks, aasbeck\}@vt.edu.} 
\thanks{ The support of the US National Science Foundation (Grants ECCS-2030215, CNS-2107276 and Grants IIS-2014499) is gratefully acknowledged. 
}
}

\maketitle
\IEEEpeerreviewmaketitle
\begin{abstract}
This paper is the first to introduce the idea of using reconfigurable intelligent surfaces (RISs) as passive devices that measure the position and orientation of certain human body parts over time. In this paper, we investigate the possibility of utilizing the available geometric information provided by on-body RISs that reflect signals from an off-body transmitter to an off-body receiver for stroke rehabilitation. More specifically, we investigate the possibility of using on-body RISs to estimate the location information over time of upper limbs that may have been impaired due to stroke. This location information can help medical professionals to estimate the possibly time varying pose and obtain progress on the rehabilitation of the upper limbs. Our analysis is focused on two scenarios: i) after assessment exercises for stroke rehabilitation when the upper limbs are resting at predefined points in the rehabilitation center, and ii) during the assessment exercises. In the first scenario, we explore the possibility of upper limb orientation estimation by deriving the Fisher information matrix (FIM) under near-field and far-field propagation conditions. It is noteworthy that the FIM quantifies how accurately we can estimate location information from a signal, and any subsequent algorithm is bounded by a function of the FIM. Coming to our propagation assumptions, the difference between the near-field and far-field regimes lies in the curvature of the wavefront. In the near-field, a receiver experiences a spherical wavefront, whereas in the far-field, the wavefront is approximately linear. The threshold to be within the near-field can be on the order of $10 \text{ m}.$  Our analysis indicates that while the upper limb orientation can be estimated when the receiver is in the near-field of the passive RIS, this orientation cannot be calculated in the far-field. In the second scenario, we present a lower bound on the achievable accuracy for the estimation of the upper limbs' location in the near-field propagation regime.  The accuracy provided by the FIM-based analysis is on the order of $0.01 \text{ rad}$ and $1 \text{ cm}$ for orientation and position of the upper limbs, respectively. This accuracy can be higher than that obtained from inertial measurement units (IMUs).  The accuracy values presented are not specific to any algorithm. Instead, the accuracy values obtained through the FIM are a benchmark for any future limb location estimation algorithm. More specifically, no limb location estimation algorithm can provide more accurate values than those obtained through the FIM. 

\end{abstract}

\section{Introduction}
The wireless channel between a transmitter and a receiver is usually considered random and uncontrollable.  This seemingly random and uncontrollable wireless channel is often treated as a nuisance that needs to be estimated and mitigated. However, the emerging idea of a reconfigurable intelligent surface (RIS) has challenged this traditional view of the wireless channel. An RIS is a collection of software-controlled subwavelength metasurfaces that perform desired transformations on incoming signals thereby controlling the propagation environment to some extent. This ability to control the incoming signal has led to several works investigating the suitability of RISs for localization \cite{10113892,emenonye2022fundamentals,9781656,9729782,8264743, 9508872,9625826,9782100,9528041,9774917}. Among other things, these works have quantified the available geometric information in signals reflected by RISs to a receiver. While this geometric information could also be used in emerging healthcare applications and wireless body area networks (WBANs), this connection has not been made yet, which has inspired this paper. 

The basic idea that we develop in this paper is to treat RISs as passive surfaces that can be placed on certain human body parts. By estimating the position and orientation of the passive RIS sensor over time, we can provide location information about the body parts on which they are placed. For concreteness, we consider an at-home rehabilitation center for stroke survivors in which RISs are considered passive reflecting devices (placed on upper limbs) that can be used to obtain location information over time about the upper limbs that may move abnormally due to stroke. This location information can help medical professionals to estimate the possibly time varying pose and obtain progress on the rehabilitation of the upper limbs. Note that this work is particularly important due to the increased life expectancy in the United States and the developed world \cite{sarker2022capturing}. Since older adults are susceptible to strokes and other medical ailments, it is very beneficial to assist this population through at-home rehabilitation schemes.

Even though the commonality of stroke mandates urgency for at-home rehabilitation, this style of healthcare is still not practical due to the amount of data required for its appropriate implementation. Note that most stroke patients suffer from hemiplegia \cite{9579464}.  Hemiplegia results in partial paralysis of one side of the body. This paralysis affects the arms, legs, and facial muscles. Hence, to address the limitations inspired by the lack of data, especially about the paralyzed arms\footnote{ It is essential to note that this use of RISs can be extended to collect more information about other body parts.}, we propose using RISs, in addition to other sensors, to collect more data.  It is essential to note that we do not offer to replace other data collection devices,  such as inertial measurement units (IMU), but we propose to use the presence of wireless signals to collect more data. Hence, the practical role of RISs in stroke rehabilitation is to collect more data. 

Also, the estimation accuracy derived from data in the wireless signal presented in this paper is not specific to any algorithm; instead, the accuracy in this paper is the highest possible accuracy that any limb location estimation algorithm can attain.    

\color{black}
\begin{figure*}[t!]
\centering
\includegraphics[width=\linewidth]{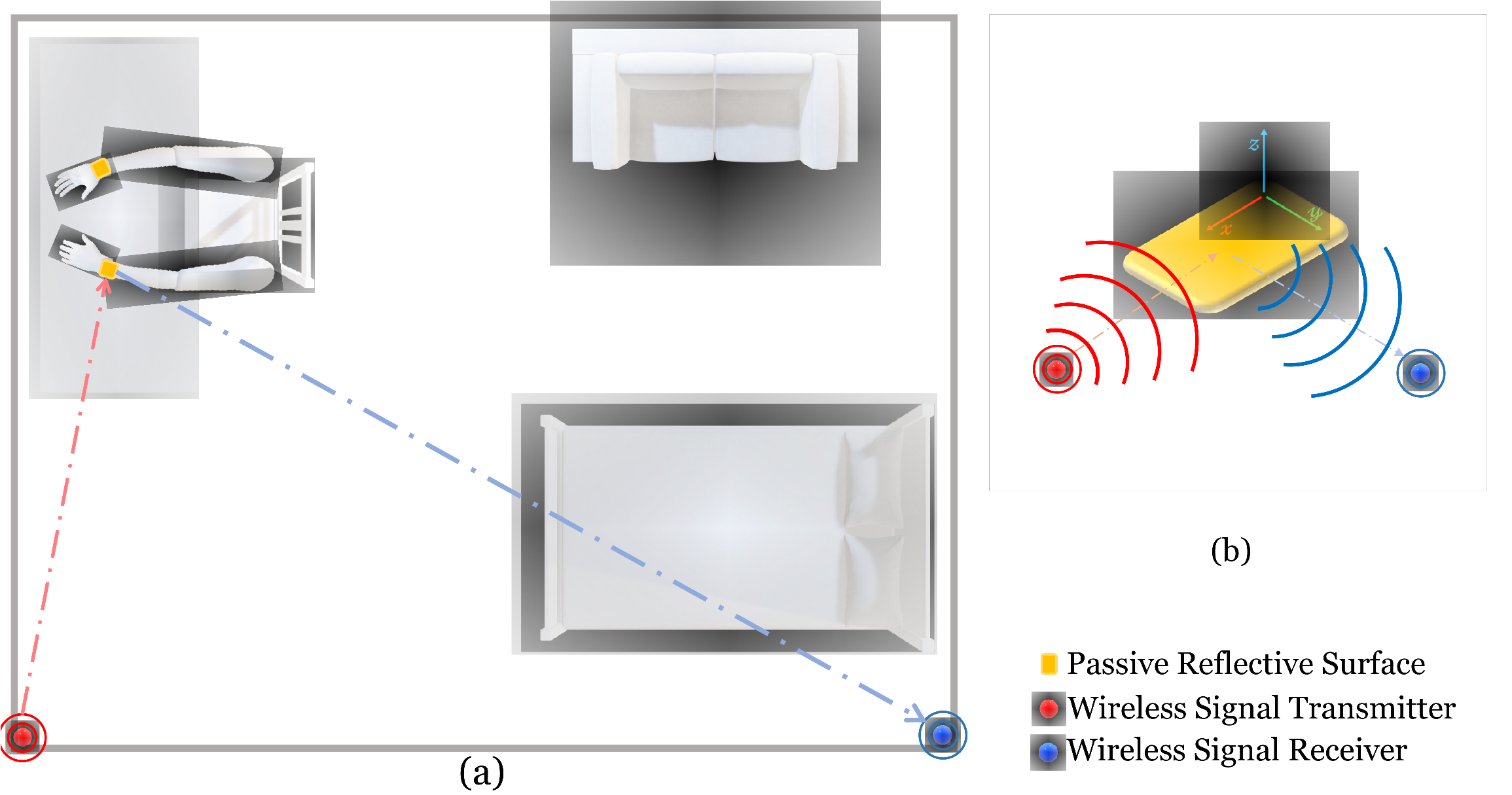}
\caption{Overview of wireless strategy for the first scenario (when the upper limbs are at rest)  (a)   transmitter (red) and receiver (blue) are installed around the home. The patient wears a small piece of passive RIS on his wrist. (b) The passive RIS is worn on the hand.  We must estimate the orientation information in the near-field propagation regime (blue).} 
\label{fig:overall_concept_figure_3}
\end{figure*}
\begin{figure*}[t!]
\centering
\includegraphics[width=\linewidth]{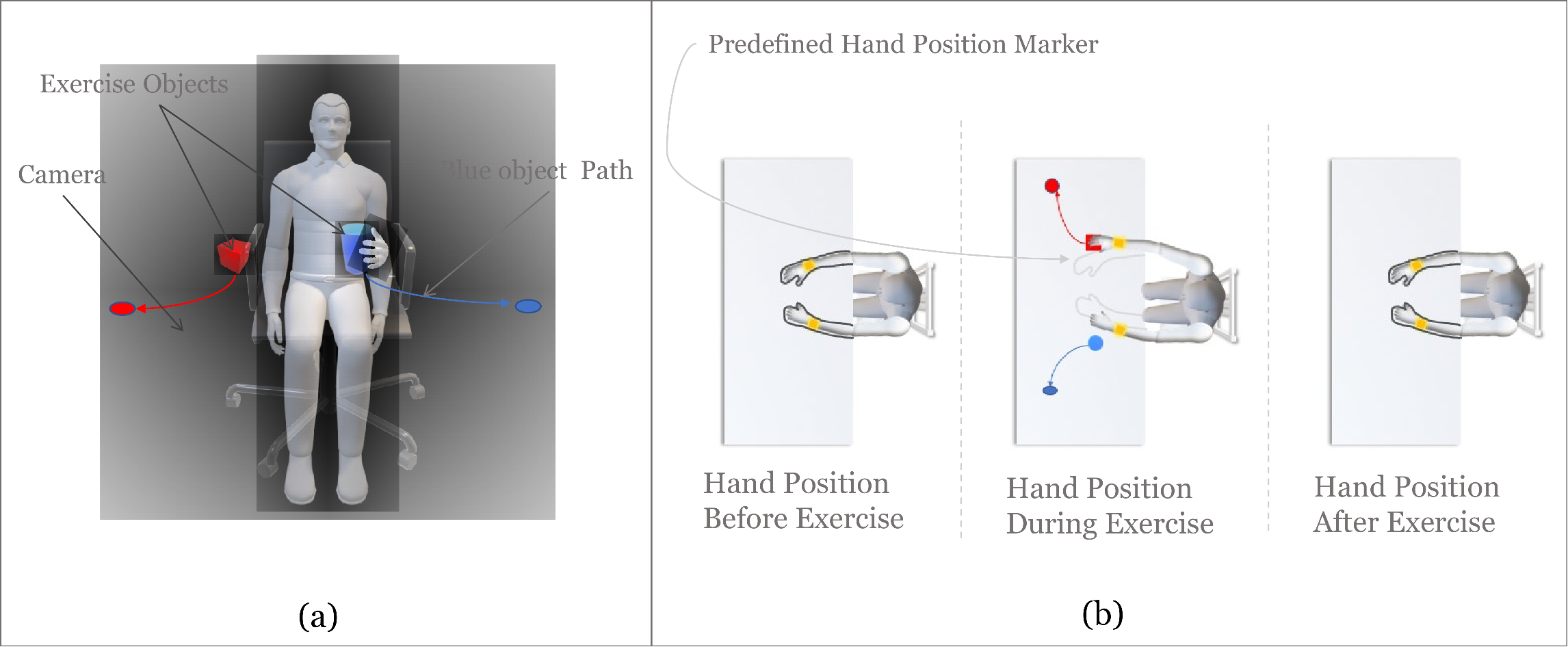}
\caption{Wireless strategy during the kinematic assessment exercises. (a) Sample setup of the Rehabilitation Center. The patient performs different exercises while sitting in a chair. Such as moving different objects from one place to another on the table. A camera is placed in front of the patient to record and monitor the performance of the exercises. (b) Scenarios: Hand position before, during, and after exercises. There is a predefined marker on the table. Before any exercises, the patient's hand position should be inside the marker. During exercises, the hand position will change based on the task. Finally, after each exercise, the patient will bring back their hand inside the marker.} 
\label{fig:excersise_sample}
\end{figure*}

\subsection{Related Works}
This paper is related to the following three research directions: i) smart health applications and WBANs, ii) pose reconstruction, and iii) analysis of the available geometric information in RIS-aided wireless systems. The relevant works in each of these areas are summarized next.
\subsubsection{Smart health applications and WBANs}
Smart health includes the research area of deploying distributed sensors in at-home rehabilitation centers for assisted living. The sensors can collect healthcare-related data and upload them to remote centers where medical professionals can analyze them. The challenges in smart health start at the data collection stage and exist at all levels. These challenges range from designing low-cost sensors for data collection, data security and privacy, efficient communication protocols, data storage, and data access management. 

In \cite{9184069,9181524,9181534}, the issue of spectral and energy efficiency is tackled for smart health networks. Authors in \cite{9184069} employ a differential chaos shift keying to tackle the limited battery level of several sensors in rehabilitation centers. In \cite{9181524}, sparse vector coding non-orthogonal multiple access (SVC-NOMA) is employed to improve spectral efficiency, while in \cite{9181534},  a game theoretic framework is used to improve data rate and spectral efficiency simultaneously. 

Generally, WBANs refer to the network of nodes implanted on a human body or close to a human body, usually monitoring the vital signs and orientation of various human body parts, subsequently providing healthcare-related data that assists medical professionals in evaluating a patient's recovery. Due to the nature of wireless propagation around the human body, WBANs have unique challenges, and several works have investigated potential solutions to these challenges. In \cite{7014261}, game theory is applied to ensure fairness in scheduling WBAN sensors while meeting the sensors' quality of service (QoS) requirements. Authors in \cite{9444832} combine sleep scheduling with energy harvesting to prolong the lifetime of WBAN sensors. In \cite{8580600}, a priority and delay-aware scheduling algorithm is investigated.  Authors in \cite{9208756} employ a Markov model to detect anomalies in order to reduce the risk of malicious attacks. The authors in \cite{9762319} optimize the RIS phase shifts of a reinforcement learning-based WBAN framework to improve learning efficiency and ensure secure data offloading. 

In \cite{7463038}, a game theoretic approach is proposed to guarantee energy efficiency and minimize the end-to-end packet delay. In \cite{9658230}, non-coherent modulation, distributed reception, and supervised learning is used to overcome the issue of outdated channel state information in WBANs. In \cite{9023978}, the  $2.4 \text{ GHz}$ and $60 \text{ GHz}$ frequency bands are investigated for WBANs; the $60 \text{ GHz}$ band is shown to be better under a variety of conditions. In \cite{8663302}, the medium access control layer is optimized to reduce latency for WBANs. In \cite{8981867}, the transmit power and encoding rate of WBAN sensors are optimized to reduce energy efficiency. In \cite{7997709}, the round trip delay is derived, and the effect of certain parameters in the MAC layer on the derived delay expression is investigated. With this plethora of works on WBANs, quite surprisingly, only one work has investigated the use RISs in WBANs  \cite{9762319}. Our paper considers the case in which RISs are passive sensors placed on the human body and reflect wireless signals to wireless access points (APs). 

\subsubsection{Pose reconstruction}
Human pose reconstruction involves localizing various body parts, usually through video data \cite{5459291,4270338,937589,1238424,1315230,1467294,6682899}, and inertia measurement units (IMU) \cite{poulose2019indoor,roetenberg2006inertial,7393844}.  Pose estimation through video data can be done by learning the space of all plausible poses \cite{5459291,4270338,937589,1238424} or mapping from image features to pose space \cite{1315230,1467294,6682899}.  In \cite{5459291,4270338,937589,1238424}, the pose estimation problem is framed as a learning problem. This learning system trained in the pose domain is assisted with prior information about the human body motion \cite{5459291}, and shape \cite{4270338}. Also, the learning problem is considered when the captured video frames have non-humans \cite{937589}, and a hashing function is used to minimize the search time in \cite{1238424}. In \cite{1315230}, the $3$D pose reconstruction problem is  investigated with video-captured silhouettes as inputs to the learning algorithm.  Authors in \cite{1467294} use a discriminative learning model to pair up typical human configurations with their realistically generated $2$D silhouettes. The work done in \cite{6682899} primarily increases the database of human poses by capturing accurate $3$D human models  of the various individuals in realistic settings under various effects such as occlusion. While the frameworks in \cite{5459291,4270338,937589,1238424,1315230,1467294,6682899} are very appealing, they require a video camera which may not always be realistic. Moreover, certain human body parts can be easily occluded during video capture. To solve this problem, IMUs have been investigated for pose reconstruction. In \cite{poulose2019indoor}, inertia measurements obtained from smartphones are used to position a human body.  Authors in \cite{roetenberg2006inertial} use IMUs to provide a full pose reconstruction for the human body; while accurate, the framework requires at least $17$ sensors. To reduce the number of sensors from $17$ to $5$, the authors in \cite{7393844} develop a framework for combining IMUs with captured video data. The review of the above papers clearly shows that a purely video-based system is not sustainable. On the other hand, a system based on IMUs will suffer from error accumulation because of sensor shift, which is an inherent limitation of IMUs \cite{sarker2022capturing,poulose2019indoor}.
 Therefore, there is a need for obtaining more measurements in order to complement these existing solutions or even develop new stand-alone solutions. Inspired by this, our paper provides a framework for obtaining more measurements about certain body parts through on-body RISs that reflect signals from an off-body transmitter to an off-body receiver.
\subsubsection{Analysis of the available geometric information in RIS-aided system}
The information available about geometric parameters in the signals reflected by  RISs has been quantified through the Fisher information matrix (FIM) for localization purposes \cite{10113892,emenonye2022fundamentals,9781656,9729782,8264743, 9508872,9625826,9782100,9528041,9774917}. Authors in \cite{10113892,emenonye2022fundamentals} rigorously analyze the geometric information in the signals reflected by RISs. While only the far-field propagation regime is considered in \cite{emenonye2022fundamentals}, both the near-field and far-field propagation regimes are considered in \cite{10113892}. In \cite{9781656}, authors present a vision for utilizing the available geometric information. In that work, the effects of massive RISs apertures, RIS phase shifters optimization, and high-frequency reflections are considered. While authors in \cite{9729782} utilize the received signals to quantify  the available geometric information, authors in \cite{8264743} approach the same objective by exploiting Maxwell's equations. Authors in \cite{9508872} analyze the available geometric information and  present optimization schemes to maximize the geometric information that can be extracted. In \cite{9625826}, each of the RIS elements acts as virtual anchors, and the available geometric information is analyzed. The joint timing offset correction and the determination of the available geometric information are considered in \cite{9782100}. Finally, the available information is analyzed under the case of non-stationarity of RISs \cite{9528041} and receiver non-stationarity \cite{9774917}. In this paper, amongst other contributions, we use the Fisher information matrix commonly used in RIS-aided localization literature to determine under what conditions the orientation of the upper limbs can be estimated when using passive RISs as on-body sensors.
\subsection{Contributions}
This paper is the first to introduce the idea of RISs as passive devices that measure the position and orientation of certain human body parts over time. To introduce this idea, this paper investigates the location information available in the signal at a receiver after reflections from an on-body RIS acting as a passive sensor for stroke rehabilitation. This location information helps medical professionals to estimate the possibly time varying pose and obtain progress on the rehabilitation of the upper limbs. We consider the analysis under two scenarios: i) after assessment exercises for stroke rehabilitation when the upper limbs are resting at predefined points and ii) during the assessment exercises. Under this framework, our main contributions are
\subsubsection{Introducing RISs as passive position and orientation measuring devices} We leverage the presence of two transceivers in a smart health setup to collect more information concerning the impaired upper limbs. 

 In addition to other on-body sensors already worn by the patient, we propose using passive RISs as sensors worn on both upper limbs. The analysis of the other on-body sensors, which typically consist of IMU units, is not the focus of this paper. For a detailed investigation of the position and orientation accuracy provided by IMU sensors, see \cite{sarker2022capturing}.
\color{black}
These passive RISs reflect the wireless signal from an off-body transmitter to an off-body receiver. These reflected signals contain location information useful for assessing the impaired upper limb. This location information helps medical professionals to estimate pose and obtain progress on the rehabilitation of the upper limbs.

\subsubsection{Derivations of the Fisher information for RISs acting as passive WBAN sensors}
We present a derivation of the FIM for a parameter vector consisting of the RISs' location parameters when they act as passive WBAN sensors under both scenarios. For a comprehensive analysis, both the near-field and far-field propagation regimes are considered in the derivations.
\subsubsection{Estimation of the orientation of the upper limbs after assessment exercises for stroke rehabilitation when the upper limbs are at rest}
In this scenario, through an analysis of the  FIM, we present the conditions in which the orientation of the on-body RIS sensors can be estimated. More specifically, we show that the possibility of estimating the orientation of the upper limbs exists in the near-field, but this possibility does not exist in the far-field. We derive the Schur complement of the FIM under both propagation regimes to present these conditions. We then show that the Schur complement is only positive definite in the near field when there is more than one antenna on the off-body receivers. With these, we conclude that the parameter vector containing the orientation information can only be estimated when the receiver is experiencing near-field propagation conditions. We also compute the error when the RISs are used as on-body orientation measuring devices through numerical simulations in a WBAN framework. We compare this orientation error to the error of a conventional orientation measuring device (a gyroscope). Finally, through an investigation of the eigenvalues of the FIM, we also present a discussion of the stability of the orientation information in the RIS-aided WBAN.  
\subsubsection{Estimation of the location (position and orientation) of the upper limbs during the assessment exercises}
In this scenario, we present the first derivatives necessary for deriving the FIM of the location parameters. Through the FIM, we compute the lower bound on the achievable accuracy for the position and orientation of upper limbs in the near-field propagation regime. The lower bound on the position and orientation information in this scenario decreases as a function of wrist size and the number of receive antennas. We compare the positioning error obtained using the passive RIS to that obtained from IMU   and a hybrid system (IMU plus video data)\cite{7393844}. We note that for similar configurations of wrist size and number of receive antennas, the lower bound on the orientation in the second scenario is much worse than the lower bound in the first scenario.
\section{Proposed Solution}  
We consider a realistic scenario where a patient's home is surveyed and partitioned such that there is a partition for the kinematics assessment of the upper limb, this partition is referred to as a rehabilitation center \cite{sarker2022capturing}. This rehabilitation center consists of a table and chair on which the patient will perform the activities needed to assess the functionality or lack of functionality of the impaired upper limb. In addition, the rehabilitation center consists of a personal computer that provides instructions to the patient. These instructions guide the patient through activities needed for the kinematics assessment. The center also includes cameras placed on tripods used to record the activities (providing visual data for analyzing the functionality of the upper limb) and on-body sensors for recording data that can provide information about the orientation and movement of the upper limb. The setup is shown in Figs. \ref{fig:overall_concept_figure_3} and \ref{fig:excersise_sample}. The table and chair are constrained to specific positions to allow for consistency in the data collected during the activities.

Moreover, after each kinematic assessment exercise, the hand is returned to an exact predefined position on the table. For patients with extreme hemiplegia, such that this is not possible, a region could be predefined. 
\color{black}
Hence, the table is marked to ensure that the patient's hand is positioned correctly. The rehabilitation center also contains wireless transceivers for connecting the patient's home to a licensed medical center; through these transceivers, the collected data is uploaded to a remote database, where medical professionals can analyze the data and provide feedback to the patient. Our contribution leverages the presence of two transceivers in the rehabilitation center to collect more information concerning the impaired upper limb. 

 In addition to other on-body sensors already worn by the patient, we propose using passive RISs as sensors worn on both upper limbs. The analysis of the other on-body sensors, which typically consist of IMU units, is not the focus of this paper. For a detailed investigation of the position and orientation accuracy provided by IMU sensors, see \cite{sarker2022capturing}. The wireless signal reflected from an off-body transmitter to an off-body receiver contains valuable location information for assessing the impaired upper limb. In addition, the location information can assist in determining the progress of the rehabilitation of the upper limbs.  \color{black}We consider the analysis under two scenarios: i)
after assessment exercises for stroke rehabilitation when the
upper limbs are resting at the predefined points (shown in Fig. \ref{fig:overall_concept_figure_3}) and ii) during
the assessment exercises (shown in Fig. \ref{fig:excersise_sample}). In the first scenario, the hand rests on the table at an exact predefined position. Hence, the position of the upper limb is known. In this scenario, we investigate the orientation information available about the upper limbs. In the second scenario, the position and orientation of the upper limb are unknown during the kinematic assessment exercises. Hence, both parameters are investigated.

\subsection{Upper Limb Orientation Information in Spherical Wavefront}
\begin{figure}[htb!]
\centering
{
\includegraphics[width=\linewidth]{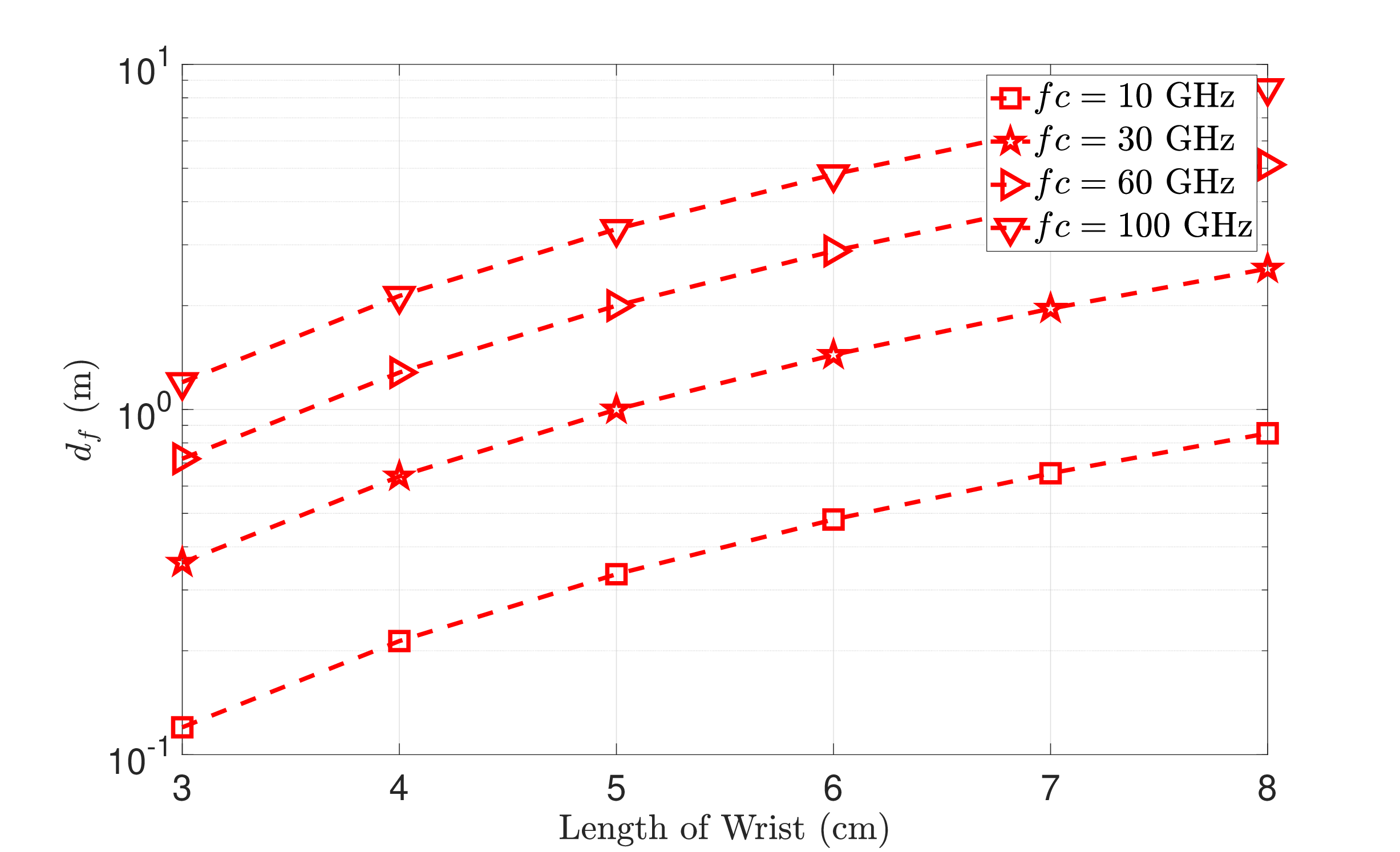}
\caption{Starting point of the Fraunhofer region for different frequencies and
different wrist sizes.}
\label{fig:Results/df_L_R}}
\end{figure}

As we will more rigorously discuss in Section \ref{section:avail_infor}, obtaining orientation information about the impaired upper limb from the signals reflected by the passive RISs is based on the availability of substantial wavefront curvature at the receiver, which is available only in the near-field propagation regime. The Fraunhofer distance $d_{\mathrm{f}}=2 D^2 / \lambda$ specifies the boundary between the near-field and far-field propagation regimes    with $\lambda$ indicating  the operating wavelength and  $D$  the diameter of the RIS \cite{9335528}. In the far-field propagation regime, the receiving antenna array experiences the wavefront as a planar wave. In contrast, the receiving antenna array experiences the wavefront as a spherical wave in the near-field propagation regime.  From the Fraunhofer distance, the availability of substantial wavefront curvature depends on the size of the RIS. However, due to human wrist size, these sensors are size-limited, and they can only have practical sizes varying from $3 \text{ cm} - 8 \text{ cm}$ in breadth. This 
 leads to small near-field propagation regions. Fig. \ref{fig:Results/df_L_R} shows the Fraunhofer distance as a function of the width of human limbs. In this Fig., the operating frequency is kept constant, and a square-shaped passive RIS is considered.
At the lowest considered frequencies, $f_c  = 10$ GHz, to get the required wavefront curvature at the receiver, the maximum distance of the receiver from the RIS sensor must be less than $0.1$ m and $0.85$ m for am RIS sensor of dimensions $3 \text{ cm} \times 3 \text{ cm}$ and $8 \text{ cm} \times 8 \text{ cm}$, respectively. At the highest considered frequencies, $f_c  = 100$ GHz, to get the required wavefront curvature at the receiver, the maximum distance of the receiver from the RIS sensor must be less than $1.2$ m and $8.5$ m for an RIS sensor of dimensions $3 \text{ cm} \times 3 \text{ cm}$ and $8 \text{ cm} \times 8 \text{ cm}$, respectively.
\subsection{Wireless Setup}
To achieve the wireless enabled assessment of the impaired upper limbs, we consider a single antenna transmitter\footnote{We assumed a single-antenna transmitter to account for the possibility of low-cost transceivers, which are still fairly common in many households. This assumption can of course be relaxed at the expense of a slightly higher notational complexity. However, the performance trends and key insights will remain the same.}, two RIS sensors with $N_{{R}}^{[{m}]}$ reflecting elements at the ${m}^{\text {th}}$ RIS sensor where ${m} \in  \mathcal{M}_1 = \{1, 2\}$,  and  a receiver with $N_{{U}}$ antennas. 

It is important to note that this number of RISs selected does not limit the generality of the proposed framework. In fact, the number of RISs can be arbitrary. The two RIS case is considered in this paper for the ease of mathematical exposition. 
\color{black}
The transmitter is located at $\bm{p}_B = [0,0,4]^{\mathrm{T}}$ and serves as the global reference for the coordinate system, which contains the location of the transceivers and the RIS sensors. The $m^{\text{th}}$ RIS sensor is located at $\bm{p}^{[m]}_{R}$, with its $r^{\text{th}}$ element located at  $\bm{p}^{[m]}_{r} = \bm{p}^{[m]}_{R} + \bm{s}^{[m]}_{r}$. The point, $\bm{s}^{[m]}_{r}$, describes the position of the $r^{\text{th}}$ on the $m^{\text{th}}$ RIS sensor with respect to $\bm{p}^{[m]}_{R}$  and can be expressed as  ${\bm{s}}_{{r}}^{[m]} = \bm{Q}_{R}^{[m]}\Tilde{\bm{s}}_{{r}}^{[m]} $, where $\bm{Q}_{R}^{[m]}$ is a $3$D rotation matrix completely specified by its angles $\bm{\Phi}_{R}^{[m]}$ \cite{lavalle2006planning}. The point $\Tilde{\bm{s}}_{{r}}^{[m]} $ describes the location of the $r^{\text{th}}$ element, if there is no orientation difference between the $m^{\text{th}}$ sensor and the global reference axis (axis of the transmitter). The receiver is located at $\bm{p}_{U}$ and its $u^{\text{th}}$ antenna is located at $\bm{p}^{}_{u} = \bm{p}^{}_{U} + \bm{s}^{}_{u}$.  The point $\bm{s}^{}_{u}$ describes the position of $u^{\text{th}}$ antenna with respect to $\bm{p}^{}_{U}$, and the receiver is placed such that its axes align perfectly with the axis of the transmitter, i.e., $\bm{s}^{}_{u} = \Tilde{\bm{s}}^{}_{u}$. The location of the ${m}^{\text {th}}$ RIS sensor can be expressed as a function of the location of the transmitter
\begin{equation}
\label{equ:pb_pr_m}
   \begin{aligned}
\bm{p}_{{R}}^{[m]} &= {\bm{p}}_{{B}}     +d_{\bm{p}_{B} \bm{p}_{R}}^{[m]} \bm{\Delta}_{\bm{p}_{B} \bm{p}_{R}}^{[m]},
    \end{aligned}
\end{equation}
where  $d_{\bm{p}_{B} \bm{p}_{R}}^{[m]}$ is the distance from point ${\bm{p}}_{{B}}$ to point $\bm{p}_{R}^{[m]}$ and $\bm{\Delta}_{\bm{p}_{B} \bm{p}_{R}}^{[m]}$ is the corresponding unit direction vector $\bm{\Delta}_{\bm{p}_{B} \bm{p}_{R}}^{[m]} = [\cos \phi_{\bm{p}_{B} \bm{p}_{R}}^{[m]} \sin \theta_{\bm{p}_{B} \bm{p}_{R}}^{[m]}, \sin \phi_{\bm{p}_{B} \bm{p}_{R}}^{[m]} \sin \theta_{\bm{p}_{B} \bm{p}_{R}}^{[m]}, \cos \theta_{\bm{p}_{B} \bm{p}_{R}}^{[m]}]^{\mathrm{T}}$. The location of the receiver can be expressed as a function of the $m^{\text{th}}$ RIS sensor as
\begin{equation}
\label{equ:pr_m_pu}
   \begin{aligned}
 {\bm{p}}_{{U}} &=  \bm{p}_{{R}}^{[m]}     +d_{\bm{p}_{R} \bm{p}_{U}}^{[m]} \bm{\Delta}_{\bm{p}_{R} \bm{p}_{U}}^{[m]},
    \end{aligned}
\end{equation}
and the  location of the receiver can be expressed as a function of the transmitter as
\begin{equation}
\label{equ:pb_pu}
   \begin{aligned}
\bm{p}_{{U}}&= {\bm{p}}_{{B}}     +d_{\bm{p}_{B} \bm{p}_{U}} \bm{\Delta}_{\bm{p}_{B} \bm{p}_{U}}.
    \end{aligned}
\end{equation}
where $d_{\bm{p}_{R} \bm{p}_{U}}^{[m]}$, $\bm{\Delta}_{\bm{p}_{R} \bm{p}_{U}}^{[m]}$, $d_{\bm{p}_{B} \bm{p}_{U}}$, and  $\bm{\Delta}_{\bm{p}_{B} \bm{p}_{U}}$ are defined similar to definitions provided about distances and unit vectors in (\ref{equ:pb_pr_m}).

\begin{table}[h!]
\caption{Table of the most commonly used symbols.}
\begin{tabular}
{|l|l|l|l|l|l|l|l|l|}
\hline 
Symbol used & Meaning \\
\hline 
$f_c$ & Operating frequency \\
\hline 
$\lambda$ & Operating wavelength  \\
\hline 
$c$ & Speed of light \\
\hline 
$N$ & Number of OFDM subcarriers \\
\hline 
$T$ & Number of OFDM symbols \\
\hline 
$N_0$ & Noise spectral density \\
\hline 
$N_R^{[m]}$ & Number of elements on the $m^{\text{{th}}}$ RIS  \\
\hline 
$N_U$ & Number of receive antennas \\
\hline 
$\bm{p}_{B}$ & Position of the centroid of the transmitter \\
\hline 
$\bm{p}_{R}^{[m]}$ &  Position of the centroid of the $m^{\text{{th}}}$ RIS  \\
\hline 
$\bm{p}_{U}$ & Position of the centroid of the receiver \\
\hline 
$\bm{s}_{r}$ & Position of the $r^{\text{th}}$ element on the  the $m^{\text{{th}}}$ RIS  \\ & with respect to the its centroid \\
\hline 
$\bm{s}_{u}$ & Position of the $u^{\text{th}}$ receiver's antenna  \\ & with respect to the receiver's centroid \\
\hline 
$\bm{p}_{r}^{[m]}$ &  Position of the $r^{\text{th}}$ element on the $m^{\text{{th}}}$ RIS  \\
\hline 
$\bm{p}_{u}$ & Position of the $u^{\text{th}}$ antenna on the receiver \\
\hline
$d_{\bm{p}_{G} \bm{p}_{V}}$ & Distance from point $\bm{p}_{G}$ to  $\bm{p}_{V}$ \\
\hline
$\Delta_{\bm{p}_{G} \bm{p}_{V}}$ & Unit vector pointing from point $\bm{p}_{G}$ to  $\bm{p}_{V}$ \\
\hline
$\theta_{\bm{p}_{G} \bm{p}_{V}}$ & Elevation angle in the Unit vector pointing \\ & from point $\bm{p}_{G}$ to  $\bm{p}_{V}$ \\
\hline
$\phi_{\bm{p}_{G} \bm{p}_{V}}$ & Azimuth angle in the Unit vector pointing \\ & from point $\bm{p}_{G}$ to  $\bm{p}_{V}$ \\
\hline
$\tau_{\bm{p}_{G} \bm{p}_{V}}$ & Time delay from point $\bm{p}_{G}$ to  $\bm{p}_{V}$ \\
\hline
$\bm{Q}^{[m]}$ & Matrix describing the orientation of the $m^{\text{th}}$ RIS \\
\hline
$\bm{\Phi}^{[m]}$ & Orientation angles of the $m^{\text{th}}$ RIS \\
\hline
$\bm{\Omega}_{t}^{[{m}]}$ & Reflection coefficients of the  $m^{\text{th}}$ RIS during the \\ &$t^{\text{th}}$ OFDM symbol \\
\hline
$\bm{\Gamma}_{t}^{[{m}]}$ & Reflection coefficients of the  $m^{\text{th}}$ RIS that \\ &remains constant across all OFDM symbols \\
\hline
$\bm{\gamma}_{t}^{[{m}]}$ & Reflection coefficients of the  $m^{\text{th}}$ RIS that \\ & changes from one OFDM symbol to the next\\
\hline
${\vartheta_{r}^{[m]}}$ & Phase of the $r^{\text{th}}$ element of the $m^{\text{th}}$ RIS \\
\hline
$\beta^{[{m}]} $ & Complex path gain of the $m^{\text{th}}$ RIS \\
\hline
$\bm{a}_{\bm{p}_B}^{[m]}$ & Transmit array response gain in the near-field related \\
& to the transmitter to $m^{\text{th}}$ RIS path \\
\hline
$\bm{a}_{\bm{p}_u}^{[m]}$ & Receive array response gain in the near-field related \\
& to the  $m^{\text{th}}$ RIS to receiver path \\
\hline
$\bm{a}_{BR}^{[m]}$ & Transmit array response gain in the far-field related \\
& to the  transmitter to $m^{\text{th}}$ RIS path \\
\hline
$\bm{a}_{RU}^{[m]}$ & Receive array response gain in the far-field related \\
& to the  $m^{\text{th}}$ RIS to receiver path \\
\hline
$\nabla$ & Represents the derivative\\
\hline
$\mathbf{J}_{\eta}$ & FIM of the parameter, $\eta$\\
\hline
\end{tabular}
\label{table_1}
\end{table}
\color{black}

\subsection{Transmission and Reception Model}
We consider the transmission and reception of $T$ OFDM symbols, each with $N$ subcarriers separated by $\Delta f$ subcarrier spacing. During the transmission of an OFDM symbol, the single antenna transmitter transforms an $N-$sized stream of data symbols from the frequency domain to the time domain using an inverse fast Fourier transform (IFFT). Subsequently, it adds an $N_{cp}-$sized cyclic prefix and the duration of the OFDM symbol is $ N_{cp}T_s + NT_s$, where $T_s = 1/B$ is the sampling time and $B = N \Delta f$ is the bandwidth.
At the $u^{\text{th}}$ receive antenna after the cyclic prefix is removed, an $N-$point fast Fourier transform (FFT) is used to transform the data symbols from the time to the frequency domain, and the received signal at the $n^{\text{th}}$ subcarrier during the $t^{\text{th}}$ OFDM symbol is written as
\begin{equation}
\label{equ:receive_processing_1}
\begin{aligned}
{y}_{t, {{u}}}[n] &=   \beta^{[0]}e^{-j 2 \pi f_{c} \tau_{{\bm{p}_{B}} {\bm{p}_{u}}}^{[0]}}x_{B}[n] \\
&+
  \sum_{{m} = 1}^{{M}_1} \beta^{[{m}]} \bm{a}^{[m]\mathrm{T}}_{\bm{p}_u}\bm{\Omega}_{t}^{[{m}]}\bm{a}_{\bm{p}_B}^{[m]} x_{B}[n] + {n}_{t, {{u}}}[n]
,\\
{y}_{t, {{u}}}[n] &=   \beta^{[0]}e^{-j 2 \pi f_{c} \tau_{{\bm{p}_{B}} {\bm{p}_{u}}}^{[0]}}x_{B}[n] \\ 
&+
  \sum_{{m} = 1}^{{M}_1} \gamma_{t}^{[{m}]}\beta^{[{m}]} \bm{a}^{[m]\mathrm{T}}_{\bm{p}_u}\bm{\Gamma}^{[m]} \bm{a}_{\bm{p}_B}^{[m]} x_{B}[n] + {n}_{t, {{u}}}[n]
,\\
&=   {\mu}^{}_{t,u}[n] +  {n}_{t,u}[n],
  \end{aligned}
\end{equation}
where  $\bm{a}_{\bm{p}_u}^{[m]} = [e^{-j 2 \pi f_{c} \tau_{{\bm{p}_{1}} {\bm{p}_{u}}}^{[{m}]}},\cdots, e^{-j 2 \pi f_{c} \tau_{{\bm{p}_{N_R^{[m]}}} {\bm{p}_{u}}}^{[{m}]} }]^{\mathrm{T}}$,  $\bm{a}_{\bm{p}_B}^{[m]} = [e^{-j 2 \pi f_{c} \tau_{{\bm{p}_{B}} {\bm{p}_{1}} }^{[{m}]}},\cdots, e^{-j 2 \pi f_{c} \tau_{{\bm{p}_{B}}{\bm{p}_{N_R^{[m]}}} }^{[{m}]} }]^{\mathrm{T}}$, and $x_{B}[n]$ is a pilot symbol. The reflection coefficient of the $m^{\text{th}}$ RIS sensor can be decomposed into   $\bm{\Omega}_{t}^{[{m}]} = \gamma_{t}^{[{m}]} \bm{\Gamma}^{[{m}]}
$ where $\gamma_{t}^{[{m}]}$ is a complex scalar value, $\bm{\Gamma}^{[{m}]} = \text{diag}(e^{j{\vartheta}_1^{[m]}}, e^{j{\vartheta}_2^{[m]}},\cdots,e^{j{\vartheta_{N_R^{[m]}}^{[m]}}} )$ is a diagonal matrix, and ${\vartheta_{r}^{[m]}}$ is the phase of the $r^{\text{th}}$ element of the $m^{\text{th}}$ RIS sensor. The scalar value, $\gamma_{t}^{[{m}]}$, is the fast-varying part of the reflection coefficient of the $m^{\text{th}}$ RIS sensor because it changes from one OFDM symbol to another. The diagonal matrix, $\bm{\Gamma}^{[{m}]}$, is the slow-varying part of the reflection coefficient of the $m^{\text{th}}$ RIS sensor.   The noise-free part (useful part) of the signal and the Fourier transformed thermal noise local to the UE's antenna array are represented by  $\bm{\mu}^{}_{t}[n]$ and $\bm{n}_{t}[n] \sim \mathcal{C}\mathcal{N}(0,N_0)$, respectively. The complex path gains of the LOS path and the $m^{\text{th}}$ RIS sensor path are represented by $\beta^{[0]}$ and $\beta^{[m]}$, respectively. The delay from the transmitter to the $u^{\text{th}}$ receive antenna is represented as $\tau_{{\bm{p}_{B}} {\bm{p}_{u}}}^{[0]} = d_{{\bm{p}_{B}} {\bm{p}_{u}}}^{[0]} / c$, where $d_{{\bm{p}_{B}} {\bm{p}_{u}}}^{[0]}$ and $c$ is the corresponding distance of the LOS path and the speed of light, respectively. The delays related to the $m^{\text{th}}$ RIS sensor path are defined similarly. The delay from the transmitter to the $r^{\text{th}}$ element on the $m^{\text{th}}$ RIS sensor is $\tau_{{\bm{p}_{B}}{\bm{p}_{r}} }^{[{m}]}$ and the delay from the $r^{\text{th}}$ element to the $u^{\text{th}}$ receive antenna is $\tau_{{\bm{p}_{r}} {\bm{p}_{u}}}^{[{m}]}$.

\subsection{Far-Field Approximation of the Received Signal}
The first order Taylor series expansion is used to express the distance from the $g^{\text{th}}$ element (located at $\bm{p}_g$) on the $G^{\text{th}}$ entity (located at $\bm{p}_G$) to the $v^{\text{th}}$ element (located at $\bm{p}_v$) on the $V^{\text{th}}$ entity (located at $\bm{p}_V$)  as $d_{\bm{p}_{g} \bm{p}_{v}} = d_{\bm{p}_{G}\bm{p}_V} + \Delta_{\bm{p}_{G}\bm{p}_V}^{\mathrm{T}} ({\bm{s}}_{v}- {\bm{s}}_{g}).$ Hence, the  delay is $\tau_{\bm{p}_{g} \bm{p}_{v}}$, and the corresponding phase shift can be expressed as  $e^{-j 2 \pi f_{c} \tau_{\bm{p}_{g}\bm{p}_v}}= e^{-j 2 \pi f_{c}
\tau_{\bm{p}_{G}\bm{p}_V}} e^{-j 2 \frac{\pi}{\lambda}  \Delta_{\bm{p}_{G}\bm{p}_V}^{\mathrm{T}} ({\bm{s}}_{v}- {\bm{s}}_{g})}.$ This approximation is equivalent to representing the spherical wavefront as a plane wave, an approximation that is only valid at distances greater than the Fraunhofer distance. With this approximation, the useful part of the signal received in the far-field at the $n^{\text{th}}$ during the $t^{\text{th}}$ OFDM symbol is
\begin{equation}
\label{equ:far_field_1}
\begin{aligned}
&\bm{\mu}_{t}[n] =  \beta^{[{0}]}\bm{a}_{UB}^{}  e^{-j 2 \pi f_{c}
\tau_{\bm{p}_{B}\bm{p}_U}^{[0]}}   + 
\sum_{{m} = 1}^{{M}_1} \gamma_{t}^{[{m}]}\beta^{[{m}]}  \bm{a}_{UR}^{[m]}\bm{a}_{RU}^{[m]\mathrm{H}}\   \\& \times \bm{\Gamma}^{[m]} \bm{a}_{RB}^{[m]} {x}_{B}[n]e^{-j 2 \pi f_{c} (\tau_{{\bm{p}_{B}} \bm{p}_{R}}^{[m]}   + \tau_{\bm{p}_{R} {\bm{p}_{U}}}^{[m]})},
\end{aligned}
\end{equation}
where $\bm{a}_{UB}^{} = e^{-j 2 \frac{\pi}{\lambda}  \Delta_{\bm{p}_{B}\bm{p}_U}^{\mathrm{T}} {\bm{S}}_{u}}$, $\bm{a}_{UR}^{[m]} = e^{-j 2 \frac{\pi}{\lambda}  \Delta_{\bm{p}_{R}\bm{p}_U}^{[m]\mathrm{T}} {\bm{S}}_{u}^{}}$, $\bm{a}_{RU}^{[m]} = e^{-j 2 \frac{\pi}{\lambda}  \Delta_{\bm{p}_{R}\bm{p}_U}^{[m]\mathrm{T}} {\bm{S}}_{r}^{[m]}}$, $\bm{a}_{RB}^{[m]} = e^{-j 2 \frac{\pi}{\lambda}  \Delta_{\bm{p}_{R}\bm{p}_U}^{[m]\mathrm{T}} {\bm{S}}_{r}^{[m]}}$, ${\bm{S}}_u = [{\bm{s}}_1, {\bm{s}}_2, \cdots, {\bm{s}}_{N_U} ],$ and ${\bm{S}}_r^{[m]} = [{\bm{s}}_1^{[m]}, {\bm{s}}_2^{[m]}, \cdots, {\bm{s}}_{N_R^{[m]}}^{[m]} ].$

\subsection{Fisher Information Matrix}
To investigate the possibility of estimating the orientation of the upper limb in the first scenario and to provide a lower bound on the estimation accuracy of the location of the upper limb in the second scenario, we introduce the FIM. The FIM, $\mathbf{J}_{ \bm{} \bm{\eta}}$, is related to the covariance of an unbiased estimate, $\hat{\bm{\eta}}$, through the following information inequality 
\begin{equation}
\begin{aligned}
\mathbb{E}_{\boldsymbol{\eta}}\left\{(\hat{\boldsymbol{\eta}}-\boldsymbol{\eta})(\hat{\boldsymbol{\eta}}-\boldsymbol{\eta})^{\mathrm{T}}\right\} \succeq \mathbf{J}_{ \bm{} \bm{\eta}}^{-1},
\end{aligned}
\end{equation}
and presents a lower bound of the minimum squared error of the unbiased estimate. {\em Note that the parameter vector, $\bm{\eta}$, can only be estimated, when the FIM, $\mathbf{J}_{ \bm{} \bm{\eta}}$, is positive definite \cite{horn2012matrix}.} The entry in the $v^{\text{th}}$ row and the $g^{\text{th}}$ column of the FIM is obtained as
\begin{equation}
\label{equ:observation_FIM_1}
[\mathbf{J}_{  \bm{\eta}}]_{[v,g]} = 
\frac{2}{N_{0}} \sum_{n=1}^{N}\sum_{t=1}^{T} \Re\left\{\nabla^{\mathrm{H}}_{{[\bm{\eta}]_{[v]}} } \bm{\mu}_{t}[n] \nabla_{{[\bm{\eta}]_{[g]}} } \bm{\mu}_{t}[n]\right\}.
\end{equation}
To separate the paths provided by the LOS and the RIS sensors, the following assumptions are made
\begin{equation}
\begin{aligned}
\label{assumption}
&\sum_{t = 1}^{T} \gamma_t^{[{m}]} = 0, \; \; \sum_{t = 1}^{T} \gamma_t^{[{m}] \mathrm{H}}\gamma_t^{[{m}]} = 1, \; \; \forall  \; m,  \text{\; \;and} \\ 
&\sum_{t = 1}^{T} \gamma_t^{[{m}_1] \mathrm{H}}\gamma_t^{[{m}_2]} = 0, \; \; \forall  \; m_1 \neq m_2.
\end{aligned}
\end{equation}
These constraints make the LOS paths and the paths associated with different RIS sensors separable. These constraints are easily achieved by assigning distinct columns of appropriately sized discrete Fourier transform matrices as the fast-varying part of the reflection coefficients of distinct RISs.

\section{Available Orientation Information  in the Received Signal During the First Reconstruction Scenario (While the Upper Limbs are at Rest)}
\label{section:avail_infor}
In this section, we will rigorously show that when the receiver is experiencing near-field propagation, an estimation algorithm can extract orientation information about the impaired upper limb from the signals reflected by the RIS sensors. However, an estimation algorithm can not extract this orientation information in the far-field.
Note that the rehabilitation center is designed to have consistency in the captured data. This consistency is achieved by placing the transmitter and receiver in the rehabilitation center at identical positions during each kinematic assessment session. Moreover, in this scenario, the limbs return to the exact position on the table after each kinematic assessment exercise. Hence the position of the transmitter $\bm{p}_{B}$, position of the RISs on the upper limbs ($\bm{p}_{R}^{[1]}, \bm{p}_{R}^{[2]}$), and the position of the receiver $\bm{p}_{U}$ are known. Since the orientations of the transmitter and receiver are also known, the only unknown parameters related to the LOS are $\bm{\eta}^{[0]} \triangleq\left[{\beta}_{\mathrm{R}}^{{[0]}}, {\beta}_{\mathrm{I}}^{{[0]}}  \right]$ where ${\beta}_{\mathrm{R}} \triangleq \Re\{{\beta}\}$, and ${\beta}_{\mathrm{I}} \triangleq \Im\{{\beta}\}$ are the real and imaginary parts of ${\beta}$, respectively. The unknown parameters related to the portion of the received signal that is provided by reflections from the $m^{\text{th}}$ RIS sensor is 
$$
{\bm{\eta}}^{[m]} = [\bm{\Phi}^{[m]}_{R} , \beta^{[m]}_{\mathrm{R}}, \beta^{[m]}_{\mathrm{I}}], 
$$
where $\beta^{[m]}_{\mathrm{R}} \triangleq \Re\{{\beta}\}$, and $\beta^{[m]}_{\mathrm{I}} \triangleq \Im\{{\beta}\}$ are the real and imaginary parts of $\beta^{[m]}$, respectively. The unknown parameter vector related to both the LOS, and the two RIS paths is defined as $\bm{\eta} = [\bm{\eta}^{[0]},\bm{\eta}^{[1]},\bm{\eta}^{[2]} ] \in \mathbb{R}^{12 \times 12}.$ Note that  in this  scenario the orientation of the upper limbs is only of interest  after each individual kinematic assessment exercise when the limbs of the patients are returned to the table. Without loss of generality, we assume that  $T$ OFDM symbols are received during this time interval.   With these assumptions, the FIM can be written as 
\begin{equation}
\label{equ:diag_FIM_1}
\mathbf{J}_{  \bm{\eta}} = \text{diag}\left[\mathbf{J}_{ \bm{\eta}^{[0]}}, \mathbf{J}_{  \bm{\eta}^{[1]}},\mathbf{J}_{  \bm{\eta}^{[2]}}\right],
\end{equation}
and the diagonal entries in the above equation is different under different propagation regimes. 
\subsection{FIM Entries   under the  Far-Field Propagation Regime}
To determine if the orientation of the upper limb can be estimated, we have to determine if the FIM, $\mathbf{J}_{  \bm{\eta}}$, is positive definite. Because $\mathbf{J}_{  \bm{\eta}}$ is a block diagonal matrix, its positive definiteness can be determined by analyzing its individual entries. First, we analyze the FIM related to the LOS path. The entries in the FIM, $\mathbf{J}_{ \bm{\eta}^{[0]}}$, of the parameters in the LOS paths are
\begin{equation}
\label{equ:diag_FIM_1_farField}
\mathbf{J}_{{{\beta}}_R^{[0]}} = \mathbf{J}_{{{{\beta}}_I^{[0]}}} = \norm{\bm{a}_{UB}^{}  e^{-j 2 \pi f_{c}
\tau_{\bm{p}_{B}\bm{p}_U}^{[0]}}}^2 \sum_{n = 1}^{N} |x_{B}[n]|^2 
\end{equation} and
$\mathbf{J}_{{[{{\beta}}_R^{[0]},{{\beta}}_I^{[0]}]}} = \mathbf{J}_{[{{\beta}}_I^{[0]},{{\beta}}_R^{[0]}]} = 0.$ Hence, the matrix, $\mathbf{J}_{ \bm{\eta}^{[0]}}$,  is positive definite. Since the FIM related to the LOS path is positive definite, we now analyze the FIM of the RIS paths. Because the RIS paths are identical, it is sufficient to consider only the $m^{\text{th}}$ path. We present the first derivatives 
\begin{equation}
\begin{aligned}
\label{lemma_appendix_equ:far_field_1}
&\nabla_{\bm{\Phi}_R}\bm{\mu}_{t}[n] =    \frac{j2\pi}{\lambda} \gamma_{t}^{[{m}]} \beta^{[{m}]}   \bm{a}_{UR} \bm{a}_{RU}^{\mathrm{H}}  \bm{K}_{RU} \bm{\Gamma}^{[m]} \bm{a}_{RB} \bm{a}_{BR}^{\mathrm{H}} \\
& \times x_{B}[n]  e^{-j 2 \pi f_{c}(\tau_{{\bm{p}_{B}} \bm{p}_{R}}^{[m]}   + \tau_{\bm{p}_{R} {\bm{p}_{U}}}^{[m]})} -    \frac{j2\pi}{\lambda} \gamma_{t}^{[{m}]} \beta^{[{m}]} \bm{a}_{UR} \bm{a}_{RU}^{\mathrm{H}}  \\ 
& \times  \bm{\Gamma}^{[m]} \bm{K}_{RB}  \bm{a}_{RB} \bm{a}_{BR}^{\mathrm{H}}  x_{B}[n]  e^{-j 2 \pi f_{c}(\tau_{{\bm{p}_{B}} \bm{p}_{R}}^{[m]}   + \tau_{\bm{p}_{R} {\bm{p}_{U}}}^{[m]})}.
\end{aligned}
\end{equation}
More first derivatives are
\begin{equation}
\begin{aligned}
&\label{lemma_appendix_equ:far_field_2}
\nabla_{{\beta}_R}\bm{\mu}_{t}[n] =  \gamma_{t}^{[{m}]}    \bm{a}_{UR} \bm{a}_{RU}^{\mathrm{H}}  \bm{\Gamma}^{[m]} \bm{a}_{RB} \bm{a}_{BR}^{\mathrm{H}}  x_{B}[n] \times \\ &e^{-j 2 \pi f_{c}(\tau_{{\bm{p}_{B}} \bm{p}_{R}}^{[m]}   + \tau_{\bm{p}_{R} {\bm{p}_{U}}}^{[m]})}, \\
&
\nabla_{{\beta}_I}\bm{\mu}_{t}[n] =  j\nabla_{{\beta}_R}\bm{\mu}_{t}[n],
\end{aligned}
\end{equation}
where $ \bm{K}_{RU} =   \text{diag}\left[\Delta_{{{\bm{p}}_{{R}^{}}{\bm{p}}_{{U}^{}}}}  ^{\mathrm{T}} (\nabla_{{\bm{\Phi}_R} }\bm{Q}_R\Tilde{\bm{S}_{r}})\right]$ and $  \bm{K}_{RB} =   \text{diag}\left[\Delta_{{{\bm{p}}_{{B}^{}}{\bm{p}}_{{R}^{}}}}  ^{\mathrm{T}} (\nabla_{{\bm{\Phi}_R} }\bm{Q}_R\Tilde{\bm{S}_{r}})\right]$. 
The FIM, $\mathbf{J}_{ \bm{\eta}^{[m]}}$, can be expressed as
\begin{equation}
    \begin{aligned}
\mathbf{J}_{{\bm{\eta}}^{[m]}} =\left[\begin{array}{ccccc}
\mathbf{J}_{ {\bm{\Phi}}_{R}^{[m]}} & 
\mathbf{J}_{ [{\bm{\Phi}}_{R}^{[m]},{{\beta}}_R^{[m]}]} &  \mathbf{J}_{ [{\bm{\Phi}}_{R}^{[m]},{{\beta}}_I^{[m]}]} \\
\mathbf{J}_{ [{\bm{\Phi}}_{R}^{[m]},{{\beta}}_R^{[m]}]}^{\mathrm{T}} & 
\mathbf{J}_{ {{\beta}}_R^{[m]}} &  0 \\
\mathbf{J}_{ [{\bm{\Phi}}_{R}^{[m]},{{\beta}}_I^{[m]}]}^{\mathrm{T}} & 
0 &  \mathbf{J}_{ {{\beta}}_I^{[m]}} \\
\end{array}\right],\end{aligned}
\end{equation}
where the entries are obtained by taking the first derivative of (\ref{equ:far_field_1}) and applying (\ref{equ:observation_FIM_1}). The following entries are related as $\mathbf{J}_{ {{\beta}}_R^{[m]}} = \mathbf{J}_{ {{\beta}}_I^{[m]}} $.
The Schur complement related to $\mathbf{J}_{{\bm{\eta}}^{[m]}}$ of the $m^{\text{th}}$ RIS path is
\begin{equation}
\begin{aligned}
&\mathbf{J}_{{\bm{\eta}}^{[m]}}^{\mathrm{e}} = \mathbf{J}_{ {\bm{\Phi}}_{R}^{[m]}}  - \\& \mathbf{J}_{ {{\beta}}_R^{[m]}}^{-1} [
 \mathbf{J}_{ [{\bm{\Phi}}_{R}^{[m]},{{\beta}}_R^{[m]}]}\mathbf{J}_{ [{\bm{\Phi}}_{R}^{[m]},{{\beta}}_R^{[m]}]}^{\mathrm{T}} + \mathbf{J}_{ [{\bm{\Phi}}_{R}^{[m]},{{\beta}}_I^{[m]}]}\mathbf{J}_{ [{\bm{\Phi}}_{R}^{[m]},{{\beta}}_I^{[m]}]}^{\mathrm{T}}].\end{aligned}
\end{equation}
With appropriate substitutions, the above Schur complement  is zero. More specifically, $\mathbf{J}_{{\bm{\eta}}^{[m]}}^{\mathrm{e}} = 0$. Hence,
$\mathbf{J}_{{\bm{\eta}}^{[m]}}$ is not positive definite. Hence, $\mathbf{J}_{  \bm{\eta}}$ is not invertible, and the orientation of the upper limb can not be estimated in the far-field propagation regime \cite{horn2012matrix}. 
\subsection{FIM Entries   under the Near-Field Propagation Regime}
Under the near-field propagation regime, to determine if the orientation of the upper limb can be estimated, we have to determine if the FIM, $\mathbf{J}_{  \bm{\eta}}$, is positive definite. Because $\mathbf{J}_{  \bm{\eta}}$ is a diagonal matrix, its positive definiteness can be determined by analyzing the individual entries. First, we analyze the FIM related to the LOS path. The entries in the LOS related FIM, $\mathbf{J}_{ \bm{\eta}^{[0]}}$, are
\begin{equation}
\label{equ:diag_FIM_1_nearField}
\mathbf{J}_{{{\beta}}_R^{[0]}} = \mathbf{J}_{{{{\beta}}_I^{[0]}}} = \sum_{u = 1}^{N_U}\norm{{e^{-j 2 \pi f_{c} \tau_{{\bm{p}_{B}} {\bm{p}_{u}}}^{[0]}}}}^2  \sum_{n = 1}^{N} |x_{B}[n]|^2 
\end{equation} and
$\mathbf{J}_{{[{{\beta}}_R^{[0]},{{\beta}}_I^{[0]}]}} = \mathbf{J}_{[{{\beta}}_I^{[0]},{{\beta}}_R^{[0]}]} = 0. 
$ Similar to the far-field case, the matrix $\mathbf{J}_{ \bm{\eta}^{[0]}}$ is positive definite. Since the FIM related to the LOS path is positive definite, we now analyze the FIM of the RIS paths. Also, similar to the far-field case, the RIS paths have identical channel parameters. Hence, it suffices to analyze the $m^{\text{th}}$ RIS path. For the $m^{\text{th}}$ RIS path when $N_U = 1$, the near-field and far-field propagation regimes are identical, and the orientation of the upper limb can not be estimated. To determine the possibility of estimating the orientation of the upper limb when $N_U > 1$, we drop the superscript $(\cdot)^{[m]}$ when notationally convenient. Subsequently, after obtaining the first derivatives, the FIM is 
\begin{equation}
\begin{aligned}
\mathbf{J}_{{\bm{\eta}}^{[m]}} =\left[\begin{array}{ccccc}
\mathbf{J}_{ {\bm{\Phi}}_{R}^{[m]}} & 
\mathbf{J}_{ [{\bm{\Phi}}_{R}^{[m]},{{\beta}}_R^{[m]}]} &  \mathbf{J}_{ [{\bm{\Phi}}_{R}^{[m]},{{\beta}}_I^{[m]}]} \\
\mathbf{J}_{ [{\bm{\Phi}}_{R}^{[m]},{{\beta}}_R^{[m]}]}^{\mathrm{T}} & 
\mathbf{J}_{ {{\beta}}_R^{[m]}} &  0 \\
\mathbf{J}_{ [{\bm{\Phi}}_{R}^{[m]},{{\beta}}_I^{[m]}]}^{\mathrm{T}} & 
0 &  \mathbf{J}_{ {{\beta}}_I^{[m]}} \\
\end{array}\right].\end{aligned}
\end{equation}
Defining $\bm{K}^{}(\bm{g}_g) = \text{diag}\left[\nabla_{{\bm{\Phi}}_{R} }\tau_{{\bm{p}_{1}} {\bm{g}_{g}}},  \cdots, \nabla_{{\bm{\Phi}}_{R} }\tau_{{\bm{p}_{N_R}} {\bm{g}_{g}}}\right]$, the FIMs in the above equation are written as
\begin{equation}
\begin{aligned}
&\mathbf{J}_{ {\bm{\Phi}}_{R}^{[m]}} = 2/N_0 (2\pi f_c)^2\sum_{n = 1}^{N} |x_B[n]|^2 |\beta^{[m]}|^2 \times \\
&\sum_{u = 1}^{N_U}\Re \left\{\bm{a}^{[m]\mathrm{H}}_{\bm{p}_B} \bigg[ \bm{\Gamma}^{[m] \mathrm{H}}\bm{K}^{*}(\bm{p}_u)  + \bm{K}^{\mathrm{H}}(\bm{p}_B)\bm{\Gamma}^{[m] \mathrm{H}}     \bigg]\bm{a}^{[m]}_{\bm{p}_u} \right.\\ &\bm{a}^{[m]\mathrm{T}}_{\bm{p}_u} 
\left.\bigg[ \bm{K}^{\mathrm{T}}(\bm{p}_u)\bm{\Gamma}^{[m]}  + \bm{\Gamma}^{[m] }\bm{K}^{}(\bm{p}_B)     \bigg] \bm{a}^{[m]}_{\bm{p}_B} \right\},
\end{aligned}
\end{equation}
\begin{equation}
\begin{aligned}
&\mathbf{J}_{ [{\bm{\Phi}}_{R}^{[m]},{{\beta}}_R^{[m]}]} =2/ N_0 (2 \pi f_c)\sum_{n = 1}^{N} |x_B[n]|^2 \times \\ &\sum_{u = 1}^{N_U}\Re \left\{ j\beta^{[m] \mathrm{H}} \bm{a}^{[m]\mathrm{H}}_{\bm{p}_B} \bigg[ \bm{\Gamma}^{[m] \mathrm{H}}\bm{K}^{*}(\bm{p}_u)  + \bm{K}^{\mathrm{H}}(\bm{p}_B)\bm{\Gamma}^{[m] \mathrm{H}}     \bigg] \times \right. \\ 
&\left.\bm{a}^{[m]}_{\bm{p}_u} \bm{a}^{[m]\mathrm{T}}_{\bm{p}_u}  \bm{\Gamma}^{[m]} \bm{a}^{[m]}_{\bm{p}_B} \right\},
\end{aligned}
\end{equation}
\begin{equation}
\begin{aligned}
&\mathbf{J}_{ [{\bm{\Phi}}_{R}^{[m]},{{\beta}}_I^{[m]}]} = -2/N_0 (2 \pi f_c)\sum_{n = 1}^{N} |x_B[n]|^2 \times \\ &\sum_{u = 1}^{N_U}\Re \left\{ \beta^{[m] \mathrm{H}} \bm{a}^{[m]\mathrm{H}}_{\bm{p}_B} \bigg[ \bm{\Gamma}^{[m] \mathrm{H}}\bm{K}^{*}(\bm{p}_u)  + \bm{K}^{\mathrm{H}}(\bm{p}_B)\bm{\Gamma}^{[m] \mathrm{H}}     \bigg] \times \right.\\ &\left.\bm{a}^{[m]}_{\bm{p}_u} \bm{a}^{[m]\mathrm{T}}_{\bm{p}_u}  \bm{\Gamma}^{[m]} \bm{a}^{[m]}_{\bm{p}_B} \right\},
\end{aligned}
\end{equation}
$$
 \mathbf{J}_{ {{\beta}}_R^{[m]}}= \mathbf{J}_{ {{\beta}}_I^{[m]}} = 2/N_0 \sum_{n = 1}^{N} |x_B[n]|^2 \sum_{u = 1}^{N_U} | \bm{a}^{[m]\mathrm{T}}_{\bm{p}_u}  \bm{\Gamma}^{[m]} \bm{a}^{[m]}_{\bm{p}_B}|^2,$$ and the Schur complement of $\mathbf{J}_{{\bm{\eta}}^{[m]}}$ is
\begin{equation}
\begin{aligned}
&\mathbf{J}_{{\bm{\eta}}^{[m]}}^{\mathrm{e}} = \mathbf{J}_{ {\bm{\Phi}}_{R}^{[m]}}  - \\& \mathbf{J}_{ {{\beta}}_R^{[m]}}^{-1} [
 \mathbf{J}_{ [{\bm{\Phi}}_{R}^{[m]},{{\beta}}_R^{[m]}]}\mathbf{J}_{ [{\bm{\Phi}}_{R}^{[m]},{{\beta}}_R^{[m]}]}^{\mathrm{T}} + \mathbf{J}_{ [{\bm{\Phi}}_{R}^{[m]},{{\beta}}_I^{[m]}]}\mathbf{J}_{ [{\bm{\Phi}}_{R}^{[m]},{{\beta}}_I^{[m]}]}^{\mathrm{T}}].\end{aligned}
\end{equation}
With appropriate substitutions, it can be shown that $\mathbf{J}_{{\bm{\eta}}^{[m]}}^{\mathrm{e}}  = 0$ when $N_U = 1$, and $\mathbf{J}_{{\bm{\eta}}^{[m]}}^{\mathrm{e}}  > 0$ when $N_U > 1$. Hence, $\mathbf{J}_{  \bm{\eta}}$ is not invertible when $N_U = 1$. However, when $N_U > 1$, $\mathbf{J}_{  \bm{\eta}}$ is invertible.  Hence, the possibility of estimating the orientation of the upper limb exists when both $N_U > 1$ and the receiver is in the near-field propagation regime defined by the Fraunhofer distance.
\subsection{Orientation Error Bounds}
To quantify the  orientation information available about the upper limbs, we introduce the orientation error bounds (OEB). The OEB of the $m^{\text{th}}$ RIS sensor can be obtained by
\begin{equation}
 \text{OEB}^{[m]}  =  \text{Tr}\{(\mathbf{J}_{{\bm{\eta}}^{[m]}}^{\mathrm{e}})^{-1}\}
\end{equation}
where $\mathbf{J}_{{\bm{\eta}}^{[m]}}^{\mathrm{e}}$ is the Schur complement of $\mathbf{J}_{{\bm{\eta}}^{[m]}}^{}$ and $\text{Tr}$ is the matrix trace operator. It is important to note the OEB is only meaningful in the near-field. This is because the possibility of estimating the  orientation information about the upper limbs only exists in the near-field.
\section{Available Location Information in the Received Signal During the Second Reconstruction Scenario (During the Kinematic Assessment Exercises)}
\label{section:avail_infor_pos}
In this section, we focus on the near-field propagation regime, and note that location information is essential to analyze the state of the impaired upper limbs during kinematic assessment exercises. During these exercises, the impaired upper limb is not constrained to any location on or off the table in the rehabilitation center. Hence, we present the achievable location accuracy of the impaired upper limb through the wireless signals reflected by the on-body RIS sensors and received at the off-body receiver. To present this accuracy, we derive the FIM related to the $m^{\text{th}}$ RIS sensor path. The parameterization for this path is $\bm{\kappa}^{[m]} = \left[ \bm{p}^{[m]\mathrm{T}}_{R}, \bm{\Phi}^{[m]\mathrm{T}}_{R},\beta^{[m]}_{\mathrm{R}}, \beta^{[m]}_{\mathrm{I}}
  \right]^{\mathrm{T}}$. The location parameters  in this path is  $\bm{\kappa}^{[m]}_1 = \left[ \bm{p}^{[m]\mathrm{T}}_{R}, \bm{\Phi}^{[m]\mathrm{T}}_{R}  \right]^{\mathrm{T}}$, and the nuisance parameters are collected as $\bm{\kappa}^{[m]}_2 = \left[ \beta^{[m]}_{\mathrm{R}}, \beta^{[m]}_{\mathrm{I}}  \right]^{\mathrm{T}}$.  All location parameters are collected as
$\bm{\kappa}^{} = \left[ \bm{\kappa}^{[1]}, \bm{\kappa}^{[2]} \right]^{\mathrm{T}}$. Now, the first derivatives related to the $m^{\text{th}}$ RIS sensor path   are
\begin{equation}
\begin{aligned}
\label{lemma_appendix_equ:near_position_1}
\medmath{\nabla_{\bm{p}_{R}^{[m]}}{\mu}_{t,u}[n]} &= \medmath{ (-j 2 \pi f_{c} )  \gamma_{t}^{[{m}]}\bm{a}^{[m]\mathrm{T}}_{\bm{p}_u}\bm{\Gamma}^{[m]} \bm{K}_{\bm{p}_R}^{[m]} \bm{a}_{\bm{p}_B}^{[m]} }x_{B}[n], \\
\medmath{\nabla_{\bm{\Phi}_{R}^{[m]}}{\mu}_{t,u}[n]} &= \medmath{(-j 2 \pi f_{c} )  \gamma_{t}^{[{m}]} \bm{a}^{[m]\mathrm{T}}_{\bm{p}_u}\bm{\Gamma}^{[m]} \bm{K}_{\bm{\Phi}_R}^{[m]} \bm{a}_{\bm{p}_B}^{[m]} }x_{B}[n]  , \\
 \medmath{\nabla_{\beta^{[m]}_{\mathrm{R}}}{\mu}_{t,u}[n]} &= \gamma_{t}^{[{m}]} \bm{a}^{[m]\mathrm{T}}_{\bm{p}_u}\bm{\Gamma}^{[m]} \bm{a}_{\bm{p}_B}^{[m]} x_{B}[n], \\
 \medmath{\nabla_{\beta^{[m]}_{\mathrm{I}}}{\mu}_{t,u}[n]} &= j\gamma_{t}^{[{m}]} \bm{a}^{[m]\mathrm{T}}_{\bm{p}_u}\bm{\Gamma}^{[m]} \bm{a}_{\bm{p}_B}^{[m]} x_{B}[n],
 \end{aligned}
\end{equation}
where the term concerning the derivatives related to the position  parameters is $$\bm{K}_{\bm{p}_R}^{[m]} = \text{diag}\left[\nabla_{{\bm{p}}_{R} }(\tau_{{\bm{p}_{1}} {\bm{p}_{u}}  }^{[{m}]} + \tau_{{\bm{p}_{B}} {\bm{p}_{1}}}^{[{m}]}),  \cdots, \nabla_{{\bm{p}}_{R} }(\tau_{{\bm{p}_{N_R^{}}} {\bm{p}_{u}}  }^{[{m}]} + \tau_{{\bm{p}_{B}} {\bm{p}_{N_R^{}}}}^{[{m}]})\right].$$ The term concerning the derivatives related to the orientation is $$\bm{K}_{\bm{\Phi}_R}^{[m]} = \text{diag}\left[\nabla_{{\bm{\Phi}}_{R}^{} }(\tau_{{\bm{p}_{1}} {\bm{p}_{u}}  }^{[{m}]} + \tau_{{\bm{p}_{B}} {\bm{p}_{1}}}^{[{m}]}),  \cdots, \nabla_{{\bm{\Phi}}_{R} }(\tau_{{\bm{p}_{N_R^{}}} {\bm{p}_{u}}  }^{[{m}]} + \tau_{{\bm{p}_{B}} {\bm{p}_{N_R^{}}}}^{[{m}]})\right].$$
Here, $\nabla_{\bm{p}_{R}^{[m]}} \tau_{{\bm{p}_{r}} {\bm{p}_{u}}  }^{[{m}]}  = (\bm{p}_r^{[m]} - \bm{p}_u ) / (c \times d_{{\bm{p}_{r}} {\bm{p}_{u}}  }^{[{m}]})$ and $\nabla_{\bm{p}_{R}^{[m]}} \tau_{{\bm{p}_{B}} {\bm{p}_{r}}}^{[{m}]} = (\bm{p}_r^{[m]} - \bm{p}_{B}^{}) / (c \times d_{{\bm{p}_{B}} {\bm{p}_{r}}}^{[{m}]})$.  The orientation related derivatives are  $\nabla_{\bm{\Phi}_{R}^{[m]}} \tau_{{\bm{p}_{r}} {\bm{p}_{u}}  }^{[{m}]}  = \bigg[(\bm{p}_r^{[m]} - \bm{p}_u ) \times (\nabla_{\bm{\Phi}_{R}^{[m]}}\bm{Q}_R^{[m]}) \Tilde{\bm{s}}_{r}^{[m]} \bigg] / (c \times d_{{\bm{p}_{r}} {\bm{p}_{u}}  }^{[{m}]})$ and $\nabla_{\bm{\Phi}_{R}^{[m]}} \tau_{{\bm{p}_{B}} {\bm{p}_{r}}}^{[{m}]} =  \bigg[(\bm{p}_r^{[m]} - \bm{p}_B ) \times (\nabla_{\bm{\Phi}_{R}^{[m]}}\bm{Q}_R^{[m]}) \Tilde{\bm{s}}_{r}^{[m]} \bigg] / (c \times d_{{\bm{p}_{B}} {\bm{p}_{r}}}^{[{m}]})$. The derivative, $\nabla_{\bm{\Phi}_{R}^{[m]}}\bm{Q}_R^{[m]}$, is the first derivative of the $3$D rotation matrix \cite{lavalle2006planning}. With these first derivatives, (\ref{equ:observation_FIM_1}), and (\ref{assumption}), the FIM for the channel parameters, $\bm{\kappa} = [\bm{\kappa}^{[1]},\bm{\kappa}^{[2]} ]$, is
\begin{equation}
\label{equ:pos_diag_FIM_1}
\mathbf{J}_{  \bm{\kappa}} = \text{diag}\left[ \mathbf{J}_{  \bm{\kappa}^{[1]}},\mathbf{J}_{  \bm{\kappa}^{[2]}}\right].
\end{equation}
Since, the nuisance parameters are not useful for positioning, we exclude them through the Schur's complement. The Schur's complement of the FIM, $\mathbf{J}_{  \bm{\kappa}}$ is
\begin{equation}
\label{equ:pos_diag_EFIM_1}
\mathbf{J}_{  \bm{\kappa}}^{\mathrm{e}} = \text{diag}\left[ \mathbf{J}_{  \bm{\kappa}^{[1]}}^{\mathrm{e}},\mathbf{J}_{  \bm{\kappa}^{[2]}}^{\mathrm{e}}\right],
\end{equation}
where 
\begin{equation}
\begin{aligned}
&\mathbf{J}_{{\bm{\kappa}}^{[m]}}^{\mathrm{e}} = \mathbf{J}_{  \bm{\kappa}^{[m]}_1} - \\& \mathbf{J}_{ {{\beta}}_R^{[m]}}^{-1} [
 \mathbf{J}_{ [{\bm{\kappa}}^{[m]}_1,{{\beta}}_R^{[m]}]}\mathbf{J}_{ [{\bm{\kappa}}^{[m]}_1,{{\beta}}_R^{[m]}]}^{\mathrm{T}} + \mathbf{J}_{ [{\bm{\kappa}}^{[m]}_1,{{\beta}}_I^{[m]}]}\mathbf{J}_{ [{\bm{\kappa}}^{[m]}_1,{{\beta}}_I^{[m]}]}^{\mathrm{T}}].\end{aligned}
\end{equation}
\subsection{Position and Orientation Error Bounds}
This section provides metrics to quantify the  position and orientation information available about the upper limbs during the kinematics assessment exercises. The PEB of the $m^{\text{th}}$ RIS sensor can be obtained by
\begin{equation}
 \text{PEB}^{[m]}  =  \text{Tr}\{[(\mathbf{J}_{  \bm{\kappa}^{[m]}}^{\mathrm{e}})^{-1}]_{[1:3,1:3]}\}.
\end{equation}
The OEB of the $m^{\text{th}}$ RIS sensor can be obtained by
\begin{equation}
 \text{OEB}^{[m]}  =  \text{Tr}\{[(\mathbf{J}_{  \bm{\kappa}^{[m]}}^{\mathrm{e}})^{-1}]_{[4:6,4:6]}\},
\end{equation}
where $\text{Tr}$ is the matrix trace operator. 
\section{Numerical Results}
In this section, we provide numerical results for both scenarios. 
To demonstrate these results, we investigate the position error bound (PEB), and orientation error bound (OEB) as a function of the number of receive antennas and the wrist size. We compare the PEB obtainable through the RIS sensors with the position error using IMUs and a hybrid system (IMU plus video data)\cite{7393844}. We also compare the OEB obtainable through the RIS sensors with the orientation error of an average low-cost gyroscope. The system setup consists of a single antenna transmitter located at $\bm{p}_B = [0,0,4]^{\mathrm{T}}$. All position vectors are in meters, and all orientation vectors are in radians. 

In the first scenario, the information provided by the RIS sensors is only of interest after each kinematic assessment exercise when the patient's hands are at rest. More specifically, in the first scenario, the information provided by the RIS sensors is only of interest when the upper limbs return to the designated position on the table after each kinematic assessment exercise. Based on this designation, the positions of the RIS sensors are known in the first scenario, but the positions of the RIS sensors are unknown in the second scenario. The position of the first RIS sensor is $\bm{p}^{[1]}_{R} = [2,2,4]^{\mathrm{T}}$ with the following rotation angles $\bm{\Phi}_{R}^{[1]} = [0.1,0.2, 0.1]^{\mathrm{T}}$, while the other RIS sensor is  located at $\bm{p}^{[2]}_{R}  = [2,2.3,4]^{\mathrm{T}}$, with the following rotation angles $\bm{\Phi}_{R}^{[2]} = [0.15,0.12, 0.1]^{\mathrm{T}}$. The transmitter serves as the global reference, and the orientation angles of the RIS sensors are defined with respect to the $3\text{D}$ axis of the transmitter. The operating wavelength is $3 \text{ mm}$, and the reflecting elements in both RIS sensors are separated by $1.5 \text{ mm}$. The transmit and receive antenna gains are both $2 \text{ dB}$, there are $N = 256$ subcarriers, the transmit power is $23 \text{ dB}$, and the noise spectral density (PSD) is $N_0 = -174 \text{ dBm/Hz}$.   The pathloss in the $m^{\text{th}}$ RIS path is described as $        \frac{\lambda^2 \sqrt{G_{B}} \sqrt{G_{U}}}{32 \pi \left(d_{{\bm{p}}_{{B}^{}}{\bm{p}}_{{r}}}^{[m]}\right)^{q_0 +1}
        \left(d_{ {\bm{p}}_{{r}}{\bm{p}}_{{u}^{}}^{}}^{[m]}\right)^{q_0 +1}}   $  where $q_0 = 0.285$ is the gain controlling factor, $G_B = 20 \text{ dB}$ and $G_U = 20 \text{ dB}$ \cite{10113892}.
\color{black}

Based on the operating frequency, the receiver experiences near-field propagation from an RIS sensor of dimensions $3 \text{ cm} \times 3 \text{ cm}$ when it is at a distance less than $1.2$ m.
For an RIS sensor of dimensions  $8 \text{ cm} \times 8 \text{ cm}$, the receiver experiences near-field at a distance less than $8.5$ m. The receiver is located at $\bm{p}^{}_{U} = [2,3,4]^{\mathrm{T}}$. The receiver is perfectly aligned with the transmitter such that its $3\text{D}$ orientation matrix is the identity matrix. With this receiver position, the receiver is in the near-field of the first and second RIS sensors for the minimum considered limbs dimensions of $3 \text{ cm} \times 3 \text{ cm}$. Since the RIS paths are identical, we focus on the PEB and OEB of the first RIS path.

For generality sake and to ensure that the PEB and OEB values presented are not dependent on any optimization algorithm, the values of the RIS phase shifters primarily denoted by $\bm{\Gamma}^{[{m}]}$ are randomly generated.
\color{black}
In all applicable plots, $L_R$, is used to specify the length of a side of the square-shaped passive RIS sensor\footnote{In our simulation, the PEB and OEB does not strictly decrease with increasing $L_R$ because of the random nature of the slow-varying RIS coefficients, $\bm{\Gamma}^{[{m}]}$.  }.

\subsection{Comparison to other Approaches}
We compare the OEB in the first scenario with the benchmark orientation error values of an Analog device  
(ADIS16490 gyroscope) that has been in use for $30$ minutes. We  present these orientation error values for different integration times. These orientation values are listed in \cite{AnalogDialogue}.

Again, the OEB values in the figures are benchmark values obtained from the Analog Devices manufacturer web sheet that detail the expected error values of an ADIS16490 gyroscope that has been in use for $30$ minutes \cite{AnalogDialogue}. It is important to note that the gyroscope values will deteriorate as a function of usage time due to drift. However, the OEB obtained from the wireless signal will not deteriorate due to drift. 
\color{black}

In \cite{poulose2019indoor}, an algorithm uses the data from the accelerometers, gyroscopes, and magnetometers in a smartphone to provide position estimates. The initial stage of this algorithm involves calculating the pitch and roll; this orientation information is used to perform step detection. Subsequently, step length and heading information are estimated. The heading information specifies the direction of motion and is estimated using data from the magnetometer. Finally,  the position information is calculated using step length and heading information.

In \cite{7393844}, the IMU and the video data are used to reconstruct the pose of the human body. While the pose estimation through the video data is very accurate, occlusion negatively affects the accuracy. Although complementary data obtained using IMUs are not affected by occlusion, they experience drifts in continuous operation. Hence, the authors in \cite{7393844} combine these two data sources and provide a hybrid approach. In this approach, the video data compensate for the drift in the IMU data, and the IMU data reduces the occlusion effect in the video data.

\subsection{Effect of Number of Receive Antennas on the OEB while the Upper limbs are at Rest}
\begin{figure}[htb!]
\centering
{
\includegraphics[width=\linewidth]{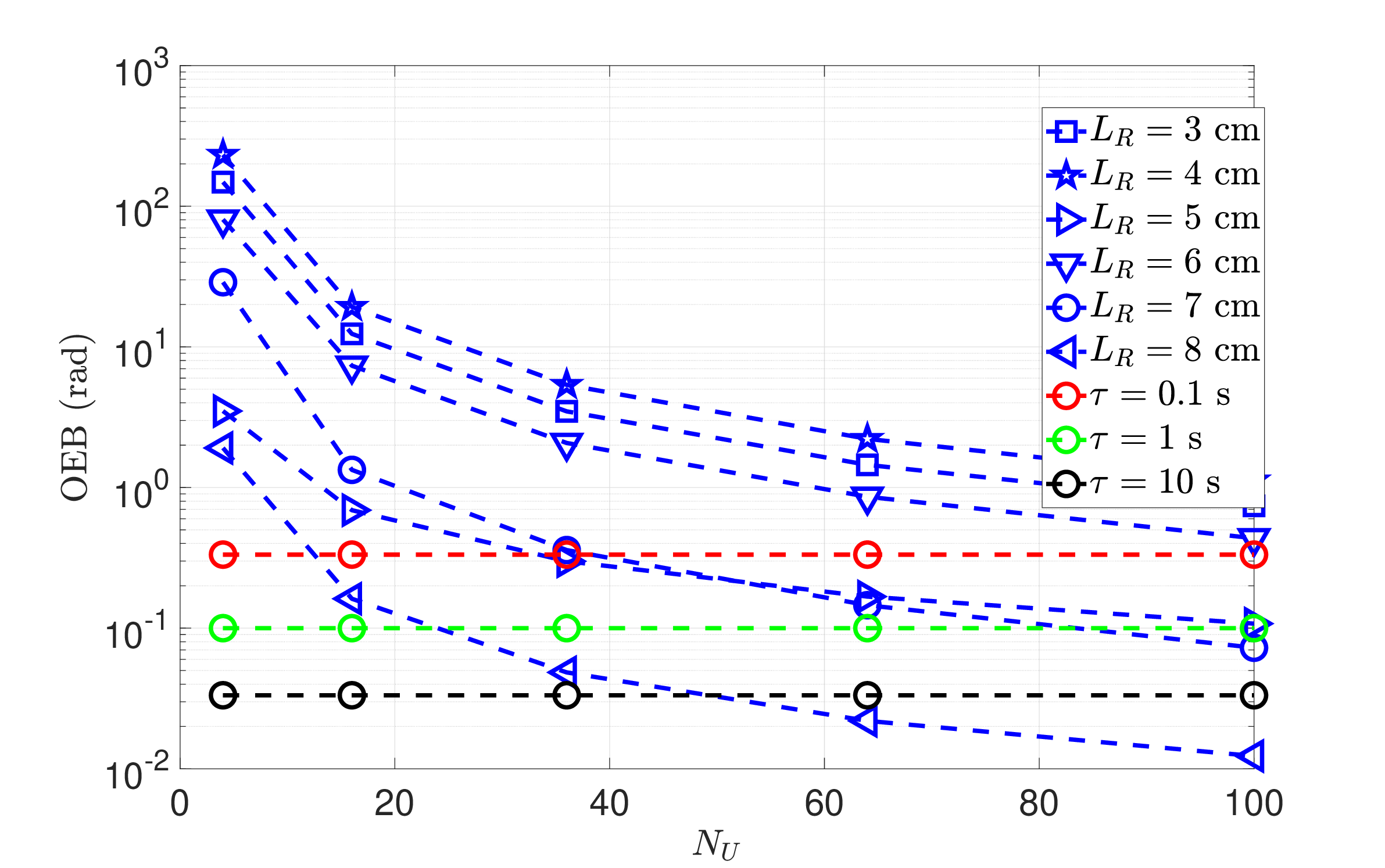}
\caption{OEB vs. the number of receive antennas. The OEB from the wireless setup is compared to the orientation error of an Analog device (ADIS16490 gyroscope). The gyroscope's error is the average error after $30 $ minutes of use,  with the variable $\tau$ representing the integration time.}
\label{fig:Results/error_1}}
\end{figure}

In Fig. \ref{fig:Results/error_1}, we investigate the available orientation information about the upper limbs as a function of both the number of receive antennas and the size of the RIS sensors. The size of the RIS sensors depends on the practical dimensions of the human limbs. 
As a comparison, we also provide the orientation error obtainable while using a cheap analog device gyroscope. The gyroscope is considered to have been in operation for $30$ minutes, and we consider different integration times for the gyroscope. The integration time of the gyroscope refers to the number of samples used to produce an orientation estimate. While the integration time of a conventional gyroscope is on the order of seconds, the time required to collect the $T$ OFDM symbols needed to provide an orientation estimate from the wireless signals depends on the signal bandwidth. This signal bandwidth at higher operating frequencies is on the order of Gigahertz. Hence the time required to collect the symbols is on the order of nanoseconds. From  Fig. \ref{fig:Results/error_1}, as the length of the RIS sensor increases, the available orientation information obtainable through the OFDM symbols received in the near-field also increases. Also, Fig. \ref{fig:Results/error_1} indicates that we can obtain more accurate orientation information as the number of receive antennas increases.

\subsection{Investigation of the  Eigenvalue of the Fisher Information while the Upper limbs are at Rest}
\begin{figure}[htb!]
\centering
{
\includegraphics[width=\linewidth]{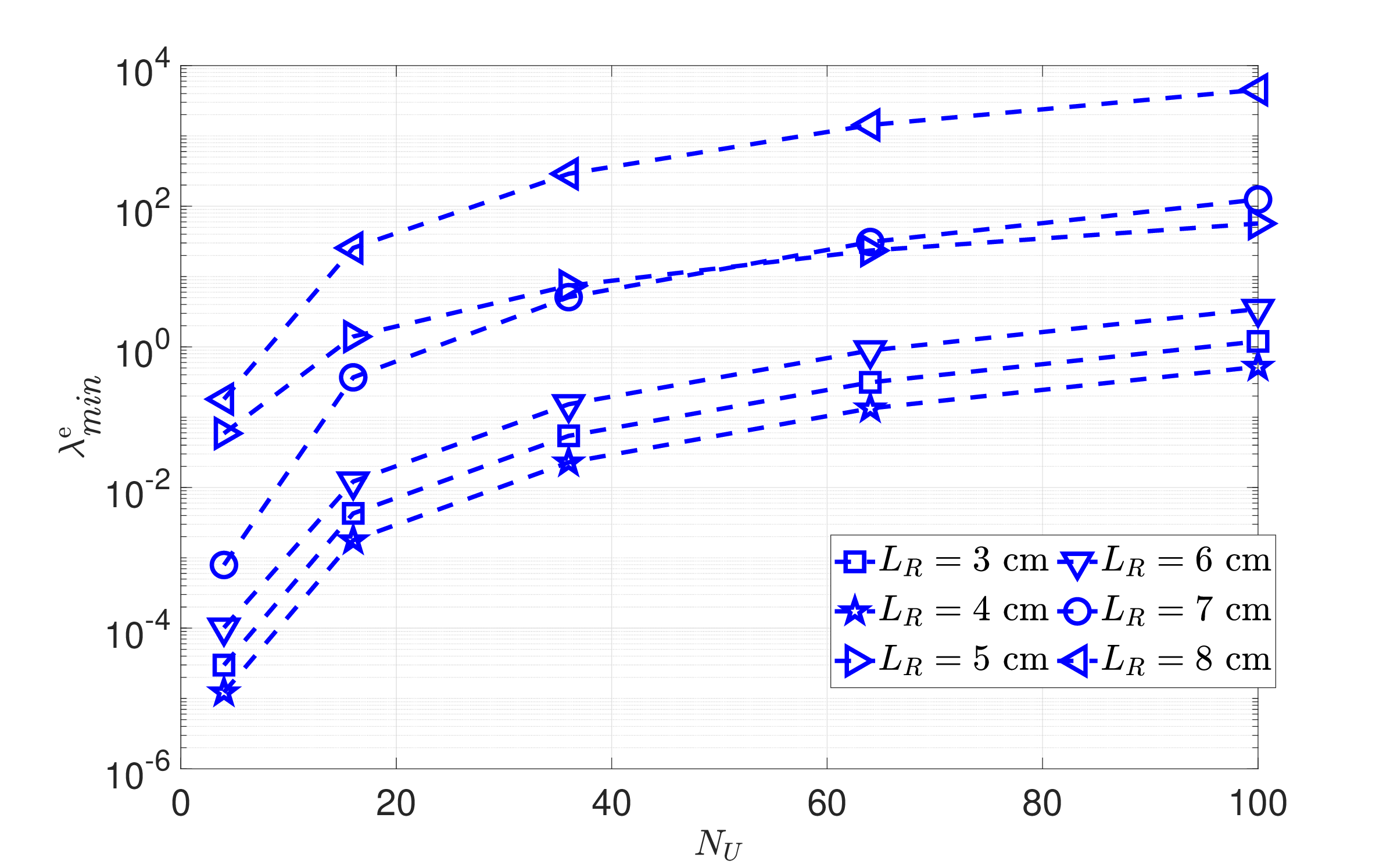}
\caption{Smallest eigenvalue vs. the number of receive antennas. }
\label{fig:Results/small}}
\end{figure}

In Fig. \ref{fig:Results/error_1}, the OEB obtainable through the received OFDM symbols is high when the RIS sensor is small and there are few numbers of receive antennas. 
\begin{figure}[htb!]
\centering
{
\includegraphics[width=\linewidth]{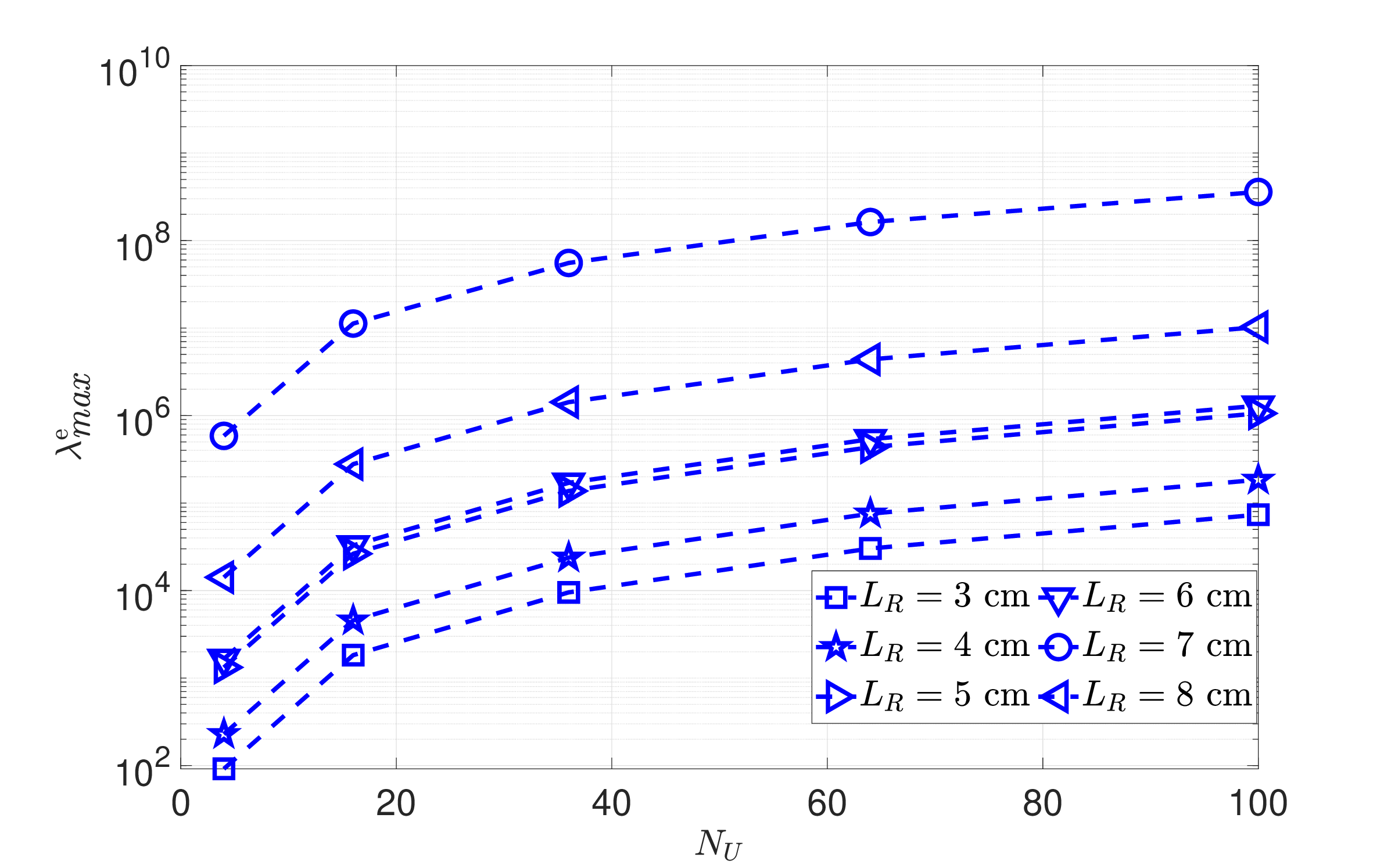}
\caption{Largest eigenvalue vs. the number of receive antennas. }
\label{fig:Results/large}}
\end{figure}
This can be explained by the instability in the orientation information under these scenarios. This instability is confirmed by investigating the structure of the Schur complement of the Fisher information matrix. More specifically, this instability is observed by investigating  the largest eigenvalue, $\lambda^{\mathrm{e}}_{\textit{max}}$, and smallest eigenvalue,  $\lambda^{\mathrm{e}}_{\textit{min}}$, of  $\mathbf{J}_{{\bm{\eta}}^{[1]}}^{\mathrm{e}}.$  These eigenvalues are shown in Fig. \ref{fig:Results/large} and Fig. \ref{fig:Results/small}. In Fig. \ref{fig:Results/small}, when the RIS sensor is small, and there are few receive antennas, the smallest eigenvalue is prohibitively small. Hence, $\mathbf{J}_{{\bm{\eta}}^{[1]}}^{\mathrm{e}}$  is almost not invertible. Because the OEB is derived by inverting $\mathbf{J}_{{\bm{\eta}}^{[1]}}^{\mathrm{e}}$, this results in high OEB values. As the number of receive antennas increases, both $\lambda^{\mathrm{e}}_{\textit{max}}$ and $\lambda^{\mathrm{e}}_{\textit{min}}$ increases and more accurate orientation  information is obtainable. 
\subsection{Position and Orientation Estimation during the Kinematic Assessment Exercises}
\begin{figure}[htb!]
\centering
{
\includegraphics[width=\linewidth]{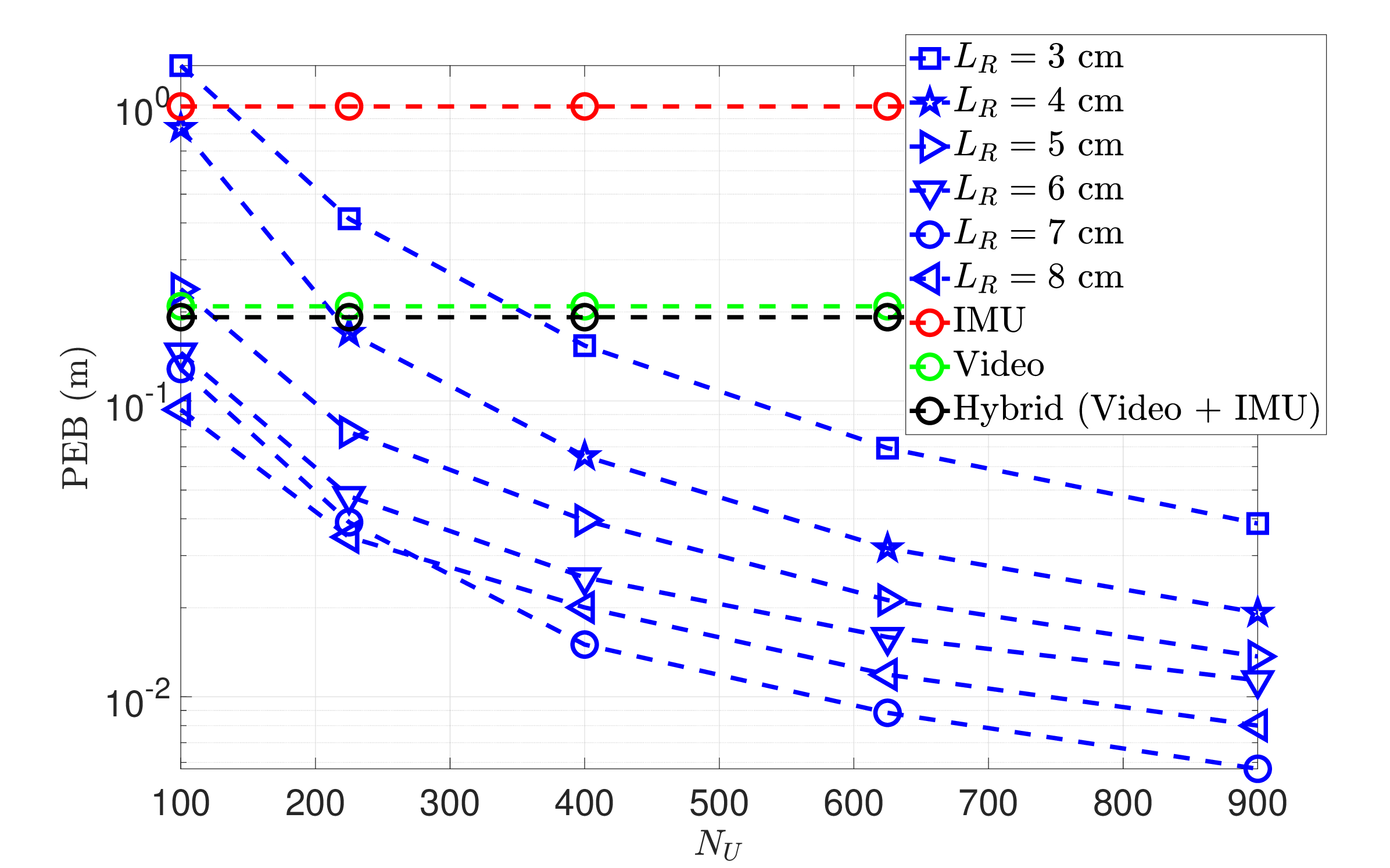}
\caption{PEB vs. the number of receive antennas. The PEB from the wireless setup is compared to the position error derived from IMU-based measurements \cite{poulose2019indoor} and from a hybrid system (IMU plus video data)\cite{7393844}.}
\label{fig:Results/errorPEB}}
\end{figure}

\begin{figure}[htb!]
\centering
{
\includegraphics[width=\linewidth]{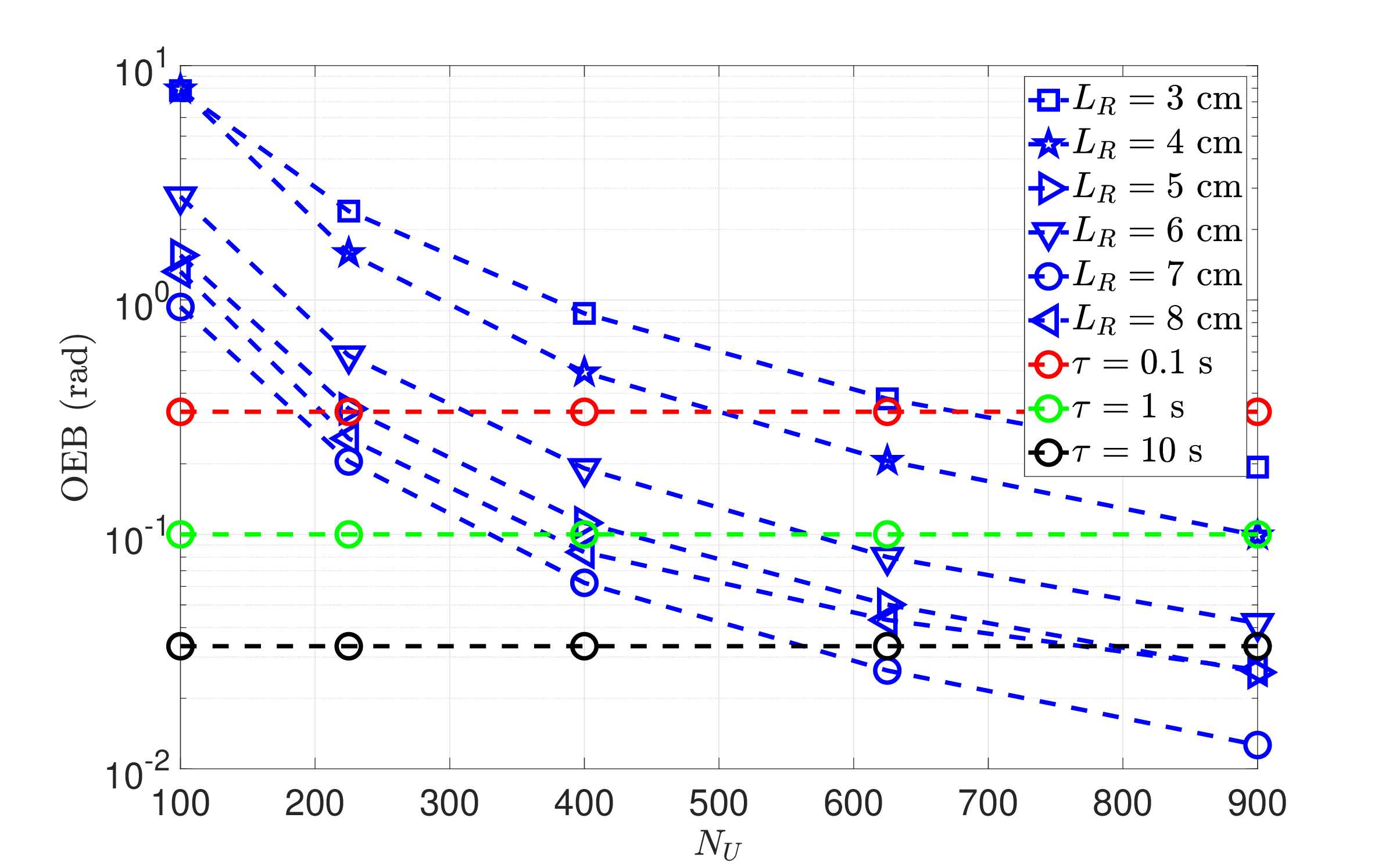}
\caption{OEB vs. the number of receive antennas. The OEB from the wireless setup is compared to the orientation error of an Analog device (ADIS16490 gyroscope). The gyroscope's error is the average error after $30 $ minutes of use,  with the variable $\tau$ representing the integration time.}
\label{fig:Results/OEBerror_1}}
\end{figure}

In this section, we present a lower bound on the estimation accuracy achievable for the position and orientation of the upper limbs through the RIS sensors during the kinematic assessment exercises. We compare this positioning error using the RIS sensor to the error achieved through IMU measurements (accelerometer, magnetometer, and gyroscope) provided by a smartphone \cite{poulose2019indoor} and through a hybrid system (IMU plus video data)\cite{7393844}. Similar to the previous section, the OEB is compared to the orientation error of an Analog device (ADIS16490 gyroscope) with different integration times. In Fig. \ref{fig:Results/errorPEB}, the PEB decreases with RIS sensor size and an increasing number of receive antennas. For most receive antenna configurations, the PEB obtained using the RIS sensor is lower than the PEB obtained using IMU measurements. The OEB derived from the RIS sensor during the kinematic assessment exercises (as shown in Fig. \ref{fig:Results/OEBerror_1}) is higher than the OEB presented in Fig. \ref{fig:Results/error_1}. This is because there are more unknown parameters during the kinematic assessment exercises.
\section{Conclusion}
This paper is the first to introduce the idea of RISs as passive devices that measure the position and orientation of certain human body parts over time. To introduce this idea, this paper has investigated passive RISs as devices that can measure the location of upper limbs over time to provide more information to medical professionals. This location information helps medical professionals to estimate the possibly time varying pose and obtain progress on the rehabilitation of the upper limbs.   The accuracy of the location information presented in this paper is not specific to any algorithm; instead, the accuracy in this paper is the highest possible accuracy that any limb location estimation algorithm can attain.    \color{black} The work in this paper is focused on two scenarios: i) while the upper limbs are at rest, and ii) during the kinematic assessment exercises for stroke rehabilitation. This investigation is carried out through a Fisher information theoretical investigation of the signals reflected from off-body transmitters by on-body passive RISs to off-body receivers. In the first scenario, we show that the orientation of the upper limb can be estimated when the receiver is experiencing near-field propagation and has more than one receive antenna. We also showed that the orientation error obtained in this RIS-enabled smart health setup can be more accurate than the orientation information provided by conventional gyroscopes. 

Also, in the first scenario, we investigated the stability of the orientation information as a function of both wrist size and the number of antennas at the receiver. The investigations indicate that the orientation information obtained is more stable for larger upper limbs. In the second scenario,  we provided a lower bound on the estimation accuracy for the position and orientation of the upper limbs while the patient is performing the kinematic assessment exercises for stroke rehabilitation. We noted that for similar configurations of wrist size and number of receive antennas, the lower bound on the orientation in the second scenario is much worse than the lower bound in the first scenario.

\bibliography{refs}
	\bibliographystyle{IEEEtran}

\end{document}